\documentclass[
reprint,
 amsmath,amssymb,
 aps,
]{revtex4-1}

\usepackage[english]{babel}
\usepackage[T1]{fontenc}
\usepackage[utf8]{inputenc}
\usepackage{bm}
\usepackage{mathtools}
\usepackage{enumerate}
\usepackage[hidelinks]{hyperref}
\usepackage{color}

\newcommand{\D}{\text d}
\newcommand{\ImI}{\text i}
\newcommand{\EuE}{\text e}

\hypersetup{
  colorlinks   = true, %Colours links instead of ugly boxes
  urlcolor     = blue, %Colour for external hyperlinks
  linkcolor    = blue, %Colour of internal links
  citecolor   = red    %Colour of citations
}

\definecolor{darkred}{rgb}{0.8, 0.0, 0.0}
\newcommand{\revcolI}[1]{#1}

\begin{document}

\title{Laplace transformed MP2 for three dimensional periodic materials using\\stochastic orbitals in the plane wave basis and correlated sampling}
\author{Tobias Sch\"afer}
\author{Benjamin Ramberger}
\author{Georg Kresse}
\affiliation{University of Vienna, Faculty of Physics and Center for Computational Materials Science, Sensengasse 8/12, A-1090 Vienna, Austria}

%\date{\today}

\begin{abstract}
We present an implementation and analysis of a stochastic high performance algorithm to calculate the correlation energy of three dimensional periodic systems in second-order M\o ller-Plesset perturbation theory (MP2). In particular we measure the scaling behavior of the sample variance and probe whether this stochastic approach is competitive if accuracies well below 1 meV per valence orbital are required, as it is necessary for calculations of adsorption, binding, or surface energies. The algorithm is based on the Laplace transformed MP2 (LTMP2) formulation in the plane wave basis. The time-dependent Hartree-Fock orbitals, appearing in the LTMP2 formulation, are stochastically rotated in the occupied and unoccupied Hilbert space. This avoids a full summation over all combinations of occupied and unoccupied orbitals, as inspired by the work of D. Neuhauser, E. Rabani, and R. Baer in J. Chem. Theory Comput. 9, 24 (2013). Additionally, correlated sampling is introduced, accelerating the statistical convergence significantly.

\end{abstract}

\maketitle

%%%%%%%%%%%%%%%%%%%%%%%%%%%%%%%%%%%%%%%%%%%%%%%%%%%%%%%%%%%%%%%%%
\section{Introduction}
%%%%%%%%%%%%%%%%%%%%%%%%%%%%%%%%%%%%%%%%%%%%%%%%%%%%%%%%%%%%%%%%%

Stochastic algorithms are considered as a promising avenue to overcome the high computational effort in wave function based methods in order to calculate the correlation energy of molecules or solids. The computation time can be reduced by introducing a statistical error. The balance between computation time and statistical error is determined by the sample variance. Hence, the sample variance is the crucial quantity to probe whether a stochastic algorithm can outperform deterministic approaches.
%One of the main benefits of stochastic approaches is the systematic balance between accuracy and computational cost. On the other hand, the sample variance determines under which circumstances such a balance outperforms deterministic approaches.

Second-order M\o ller-Plesset perturbation theory (MP2) \cite{Moeller1934} is the simplest wave function based method to calculate the correlation energy, but it already exhibits the typical problem that it is computationally fairly demanding due to the dependence on unoccupied orbitals. Therefore MP2 serves as a very suitable test candidate for new algorithms. An implementation of the canonical MP2 formulation \cite{Szabo1996} leads to a quintic scaling with the system size, whereas a quartic scaling is possible in the Green's function formulation \cite{Willow2012} and in the Laplace transformed formulation \cite{Almlof1991,Schafer2016}. This steep scaling is particularly problematic if accurate MP2 energies for three dimensional periodic materials are desired. Despite its computational complexity, MP2 is of high interest in computational chemistry and materials physics \cite{Manby2006,Casassa2008,Halo2009,Halo2009a,Erba2009,Erba2011,Casassa2012,Fabiano2009,Maschio2010,Maschio2010a,Schwerdtfeger2010,Nanda2012,Stodt2012,Goltl2012,Muller2013,DelBen2013,DelBen2014,DelBen2015,Torabi2014,Hammerschmidt2015,Kaawar2016}, as it captures most of the correlation energy and covers covalent, ionic, and van der Waals interactions.

To overcome this high computational demand, many attempts for improvements have been made for periodic systems in the recent past. Full periodic MP2 codes in the plane wave basis for three dimensional systems are available in Ref. \cite{Marsman2009, Gruneis2010, Schafer2016} where a quartic scaling was reached without approximations in \cite{Schafer2016}. Local MP2 approaches (LMP2) \cite{Pisani2005,Pisani2008,Usvyat2015} exploit the locality of the atomic orbitals, and the resolution of identity approximation (RI) \cite{Katouda2010} aims for a faster evaluation of two-electron integrals. Both RI and LMP2 are combined in Ref. \cite{Izmaylov2008,Maschio2007,Usvyat2006}. High performance codes, designed to perform the MP2 calculations on large super computers, are also discussed in Ref. \cite{DelBen2012,Schafer2016}. The very early attempts for periodic MP2 calculations should also be mentioned here \cite{Suhai1983,Sun1996}.

Also several stochastic approaches have been implemented for MP2, including real space Monte Carlo integrations of Green's functions \cite{Willow2012, Willow2014}, stochastic orbitals \cite{Neuhauser2013}, a guided stochastic energy-domain formulation \cite{Ge2014}, and a stochastic formulation of the resolution of identity \cite{Takeshita}. \revcolI{Neuhauser, Baer, and Zgid recently  published a similar algorithm as the one we use here \cite{Neuhauser2017}. Their algorithm can be even extended to self-consistent Green's functions and finite temperature.}

In this work we present a stochastic MP2 approach which is inspired by the work of Neuhauser et. al. \cite{Neuhauser2013}. Based on the unitary invariant Laplace transformed MP2 formulation (LTMP2), the time-dependent HF orbitals are rotated stochastically in the Hilbert space. The LTMP2 expression is then evaluated with these stochastic orbitals in the plane wave basis, giving a stochastic energy whose expectation value is the MP2 energy. As an extension we implemented correlated sampling which drastically speeds up the calculation. The algorithm is highly parallelized with MPI and OpenMP. One of our main objectives is to study the scaling of the variance with the system size and whether this method is competitive on systems with about 100 valence orbitals when absolute accuracies below 1 meV per valence orbital are required, like for adsorption energies, binding energies, or surface energies. If a fixed absolute statistical error, independent of the system size, is desired, the algorithm scales cubically with system size, whereas linear scaling can be observed in the case of a fixed relative statistical error (per valence orbital). The algorithm is implemented in the Vienna ab initio simulation package (VASP) \cite{Kresse1993,Kresse1999}.

The subsequent Sec. \ref{sec:theory} contains a brief repetition of the canonical and Laplace transformed MP2 formulation. Also, stochastic orbitals and correlated sampling are introduced. In Sec. \ref{sec:implementation} a comprehensive description of the implementation and a theoretical analysis of the system size scaling is presented. The benchmark calculations, where we study the competitivity of our algorithm, are provided in Sec. \ref{sec:benchmark}.

% Sage, dass der Unterschied zwischen hier und Neuhauser in Kapitel III beschrieben wird

% cite Möller and Plesset
% introduce HF

%%%%%%%%%%%%%%%%%%%%%%%%%%%%%%%%%%%%%%%%%%%%%%%%%%%%%%%%%%%%%%%%%
\section{Theory} \label{sec:theory}
%%%%%%%%%%%%%%%%%%%%%%%%%%%%%%%%%%%%%%%%%%%%%%%%%%%%%%%%%%%%%%%%%

\subsection{Laplace transformed MP2}

The text book expression of the MP2 correlation energy \cite{Szabo1996},
\begin{equation}
E^{(2)} = \frac 1 2 \sum_{ij}^{\text{occ.}} \sum_{ab}^{\text{virt.}} \frac{\langle ij | ab \rangle[\langle ab| ij\rangle - \langle ab| ji\rangle] }{\varepsilon_{i} + \varepsilon_{j} - \varepsilon_{a} - \varepsilon_{ b } } \;, \label{eq:CanMP2}
\end{equation}
is usually derived using Rayleigh-Schr\"odinger perturbation theory for the ground state energy of the many-body Hamiltonian on top of Hartree-Fock (HF). To lowest order the perturbation series yields the HF energy, $E_\text{\text{HF}} = E^{(0)} + E^{(1)}$. Hence the lowest order correction for the correlation energy is given by the MP2 energy, $E^{(2)}$. In Eq. (\ref{eq:CanMP2}), $i,j$ and $a,b$ denote the occupied and unoccupied (virtual) HF spin-orbitals, and the $\varepsilon$'s are the HF spin-orbital energies. The two-electron integrals are identified with
\begin{equation}
\langle ij | ab \rangle = \int \D^3 r \int \D^3 r' \; \frac{\varphi_i^*(\bm r) \varphi_j^*(\bm r')\varphi_a(\bm r)\varphi_b(\bm r') }{|\bm r - \bm r'|}  \;,
\end{equation}
where we have used Hartree atomic units.

It is common to divide the MP2 expression into the direct MP2 energy, $E_{\text d}^{(2)}$, containing $\langle ij | ab \rangle\langle ab| ij\rangle$  and the exchange MP2 energy, $E_{\text x}^{(2)}$, containing $-\langle ij | ab \rangle\langle ab| ji\rangle$. If closed-shell systems are considered the direct and exchange MP2 energy can be written as
\begin{align}
E_{\text d}^{(2)} &= 2 \sum_{ij}^{\text{occ.}} \sum_{ab}^{\text{virt.}} \frac{\langle ij | ab \rangle\langle ab| ij\rangle }{\varepsilon_{i} + \varepsilon_{j} - \varepsilon_{a} - \varepsilon_{ b } } \;, \\
E_{\text x}^{(2)} &= - \sum_{ij}^{\text{occ.}} \sum_{ab}^{\text{virt.}} \frac{\langle ij | ab \rangle\langle ab| ji\rangle }{\varepsilon_{i} + \varepsilon_{j} - \varepsilon_{a} - \varepsilon_{ b } } \;,
\end{align}
where the indices now only describe spatial orbitals instead of spin-orbitals.

For the sake of a compact notation we present derivations only for the exchange MP2 energy. All derivations can be applied to the direct MP2 energy in the same way.

For periodic systems the indices $i,j$ and $a,b$ have to be understood as compound indices containing the band index, the crystal wave vector, and the spin. However, for simplicity we restrict to periodic systems given by large supercells in order to avoid the extensive notation for the k-point sampling of the Brillouine zone. Hence, in this work, the Brillouine zone is sampled only by the $\bm \Gamma$-point.

As discussed by Alml\"of \cite{Almlof1991} the canonical MP2 expression (\ref{eq:CanMP2}) can be reformulated by a Laplace transform such that the orbitals do not have to be restricted to HF orbitals but any set of orbitals obtained by a unitary transformation of time-evolved HF orbitals can be used. In the first step, the energy denominator is rewritten as a Laplace transform, i.e. $1/x = -\int_0^\infty \EuE^{x\tau}\D\tau, \, x<0$, 
\begin{equation}
E_{\text x}^{(2)} =\frac 1 2 \int_0^\infty \sum_{ij}^{\text{occ.}} \sum_{ab}^{\text{virt.}} \langle ij | ab \rangle \langle ab| ji\rangle \EuE^{(\varepsilon_{i} + \varepsilon_{j} - \varepsilon_{a} - \varepsilon_{ b })\tau }\, \D\tau \;. \label{eq:LTMP2}
\end{equation}
Note that $(\varepsilon_{i} + \varepsilon_{j} - \varepsilon_{a} - \varepsilon_{ b })$ is always negative for gapped systems. In practice the integration over $\tau$ is implemented by a quadrature \cite{Haser1992, Kaltak2014}, $\int_0^\infty ... \, \D\tau \approx \sum_\tau w_\tau ...$, where $w_\tau$ is a weighting factor. According to our experience, six $\tau$-points are sufficient for all materials in this work when the $\tau$-point meshes of \cite{Kaltak2014} are employed. In the next step, time-dependent HF orbitals are defined by 
\begin{equation}
|\phi_i^\tau\rangle = \EuE^{\varepsilon_i\tau/2}|i\rangle \;,\quad |\phi_a^\tau\rangle = \EuE^{-\varepsilon_a\tau/2}|a\rangle\;, \label{eq:tdHFo},
\end{equation}
\revcolI{which relates to the square root of the Green's function $\sqrt{|G(\tau)|}$ in Ref. \onlinecite{Neuhauser2017}.} This gives rise to the formulation
\begin{multline}
E_{\text x}^{(2)} = \frac 1 2 \int_0^\infty \sum_{ij}^{\text{occ.}} \sum_{ab}^{\text{virt.}} \langle \phi_i^\tau \phi_j^\tau | \phi_a^\tau \phi_b^\tau \rangle \langle \phi_a^\tau \phi_b^\tau| \phi_j^\tau \phi_i^\tau\rangle \, \D\tau\;  , \label{eq:invLTMP2}
\end{multline}
which is invariant under unitary transformations of the time-dependent HF orbitals. This invariance can be recognized by defining a new set of unitary transformed time-dependent orbitals via
\begin{equation}
|\psi_i^\tau\rangle = \sum^{\text{occ.}}_k u_{ik}|\phi_k^\tau\rangle \;, \quad
|\psi_a^\tau\rangle = \sum^{\text{virt.}}_c v_{ac}|\phi_c^\tau\rangle \;,
\end{equation}
where $u$ and $v$ are two arbitrary unitary matrices in the occupied and unoccupied manifold, respectively. After replacing all $\phi$'s by $\psi$'s in Eq. (\ref{eq:invLTMP2}), the unitary matrices lead to Kronecker deltas that take care of the invariance:
\begin{multline}
\sum_{ij}^{\text{occ.}} \sum_{ab}^{\text{virt.}} \langle \psi_i^\tau \psi_j^\tau | \psi_a^\tau \psi_b^\tau \rangle \langle \psi_a^\tau \psi_b^\tau| \psi_j^\tau \psi_i^\tau\rangle  \\
= \sum_{{ij}\atop{k_1...k_4}}^{\text{occ.}} \sum_{{ab}\atop{c_1...c_4}}^{\text{virt.}} \langle \phi_{k_1}^\tau \phi_{k_2}^\tau | \phi_{c_1}^\tau \phi_{c_2}^\tau \rangle \langle \phi_{c_3}^\tau \phi_{c_4}^\tau| \phi_{k_3}^\tau \phi_{k_4}^\tau\rangle \\
\times \underbrace{u^{*}_{ik_1}u_{ik_4}}_{\delta_{k_1k_4}} \underbrace{u^{*}_{jk_2}u_{jk_3}}_{\delta_{k_2k_3}} \underbrace{v_{ac_1}v^{*}_{ac_3}}_{\delta_{c_1c_3}} \underbrace{v_{bc_2}v^{*}_{bc_4}}_{\delta_{c_2c_4}} \\
= \sum_{ij}^{\text{occ.}} \sum_{ab}^{\text{virt.}} \langle \phi_i^\tau \phi_j^\tau | \phi_a^\tau \phi_b^\tau \rangle \langle \phi_a^\tau \phi_b^\tau| \phi_j^\tau \phi_i^\tau\rangle \label{eq:prooveInv}
\end{multline}
 
\subsection{Laplace transformed MP2 with stochastic orbitals in the plane wave basis} \label{sec:LTMP2_with_rnd_orbtials}

When random coefficients and expectation values are used, the Kronecker deltas in Eq. (\ref{eq:prooveInv}) can be generated by yet another transformation, which, for MP2, was first published by Neuhauser et al. \cite{Neuhauser2013}. Consider a set of independent complex random coefficients $\{p_i\}$ where both the real and imaginary parts are uniformly distributed over the range $[-\sqrt{3/2},+\sqrt{3/2}]$, such that we find for the expectation values $\text E[p_i] = 0$ and $\text E[p_i^*p_i] = 1$. We can then write the Kronecker delta as an expectation value: $\delta_{ij} = \text E[p_i^*p_j]$. Note that the upright letter $\text E[...]$ stands for an expectation value, whereas energies are written by the italic letter $E$. If we plug this definition of Kronecker deltas into the third line of Eq. (\ref{eq:prooveInv}), we find that the MP2 energy can be written as an expectation value of stochastic energies $X^\tau$,
\begin{equation}
E_{\text x}^{(2)} = \frac 1 2 \int_0^\infty \text E [X^{\tau} ] \, \D\tau\;,
\end{equation}
where
\begin{equation}
X^\tau  =   \langle \kappa^\tau \lambda^\tau | \alpha^\tau \beta^\tau \rangle \langle \alpha^\tau \beta^\tau| \lambda^\tau \kappa^\tau\rangle \;, \label{eq:sample}
\end{equation}
with the stochastic orbitals
\begin{align}
&|\kappa^\tau\rangle = \sum_i^{\text{occ.}} p_i |\phi_i^\tau \rangle \;,\quad\, |\lambda^\tau\rangle = \sum_i^{\text{occ.}} q_i |\phi_i^\tau \rangle \;, \nonumber\\
&|\alpha^\tau\rangle = \sum_a^{\text{virt.}} r_a |\phi_a^\tau \rangle \;,\quad |\beta^\tau\rangle = \sum_a^{\text{virt.}} s_a |\phi_a^\tau \rangle \;. \label{eq:StochOrbit}
\end{align}
Here $\{p_i\},\{q_i\},\{r_a\}$, and $\{s_a\}$ are independent sets of uniformly distributed complex random coefficients as described above. The main idea is to generate a sufficiently large sample of the stochastic energies $X^\tau$ in order to obtain a reliable estimation for the expectation value $\text E [X^{\tau} ]$ and therefore an estimation for the MP2 energy. 

In this work, we assume that the occupied and virtual HF orbitals and energies are available through a preceding HF calculation. The HF orbitals are stored in the plane wave basis, $\{\langle \bm G| i\rangle, \langle \bm G| a\rangle\}$, where $\bm G$ is a reciprocal lattice vector. In VASP the number of lattice vectors $\bm G$ is truncated by a cutoff, $E_\text{cut}$ (\texttt{ENCUT} flag in VASP), such that $\bm G^2 / 2 < E_\text{cut}$. Also the number of orbitals (sum of occupied plus unoccupied) is limited to the same number as the number of reciprocal lattice vectors. 

In order to calculate a single stochastic energy $X^\tau$ in Eq. (\ref{eq:sample}), the stochastic orbitals are set up in the plane wave basis using (\ref{eq:StochOrbit}) and (\ref{eq:tdHFo}), e.g.
\begin{equation}
\langle \bm G |\kappa^\tau\rangle = \sum_i^{\text{occ.}} p_i \EuE^{\varepsilon_i \tau /2}\langle \bm G|i \rangle \;. \label{eq:StochOrbitPW}
\end{equation}
The two-electron integrals in (\ref{eq:sample}) are evaluated in reciprocal space as
\begin{equation}
\langle \kappa^\tau \lambda^\tau | \alpha^\tau \beta^\tau \rangle  = \frac 1 \Omega \sum_{\bm G}^{E_{\text{cut}}^{\text{aux}}} \frac{4\pi}{\bm G^2}\,  \langle \kappa^\tau | \EuE^{-\ImI\bm G \hat {\bm r}} | \alpha^\tau \rangle \langle \lambda^\tau | \EuE^{+\ImI\bm G \hat {\bm r}} | \beta^\tau \rangle  \;. \label{eq:SCH2EI} 
\end{equation}
Note that here the reciprocal lattice vectors are limited by an auxiliary cutoff, $E_{\text{cut}}^{\text{aux}}$ (\texttt{ENCUTGW} flag in VASP), which is usually equal to $\frac 2 3 E_{\text{cut}}$. Also, $\Omega$ is the volume of the system and $\langle \kappa^\tau | \EuE^{-\ImI\bm G \hat {\bm r}} | \alpha^\tau \rangle$ are so called overlap densities which are defined by
\begin{align}
\langle \kappa^\tau | \EuE^{-\ImI\bm G \hat {\bm r}} | \alpha^\tau \rangle &= \int_\Omega \D^3 r \; \langle \kappa^\tau | \bm r \rangle \langle \bm r | \alpha^\tau \rangle \, \EuE^{-\ImI\bm G {\bm r}} \nonumber\\
&= \mathcal{F}_{\bm G}[\{ \langle \kappa^\tau | \bm r \rangle \langle \bm r | \alpha^\tau \rangle  \}]   \;, \label{eq:overlap}
\end{align}
where the stochastic orbitals in real space can easily be obtained by a Fourier transform,
\begin{equation}
\{\langle \bm r | \kappa^\tau \rangle\} = \mathcal{F}^{-1}_{\bm r}[\{\langle \bm G | \kappa^\tau \rangle\}] \;.
\end{equation}
In this way, a stochastic energy $X^\tau$ can be calculated for a given $\tau$-point.

\subsection{Variance and error} \label{sec:var_error}

To estimate the expectation value of the sample for a given $\tau$-point, $\text E[X^\tau]$, the mean, $\mu^\tau_{n}$, is calculated by 
\begin{equation}
\mu_{n}^\tau = \frac 1 {n} \sum_{l=1}^{n} X^\tau_l\;,
\end{equation}
where $n$ is the number of all generated stochastic energies of this sample, since $\mu^\tau_{n} \rightarrow \text E [X^{\tau}]$ as $n \rightarrow \infty$. To measure the reliability of this estimation for finite $n$, the error of the mean is estimated via
\begin{equation}
\delta\mu^\tau_{n} = \frac{\sigma_{n}^\tau}{\sqrt{n}} \;. \label{eq:stat_error}
\end{equation}
Here $\sigma^\tau_{n}$ is an estimate for the standard deviation of the samples, obeying $\sigma^\tau_{n} \rightarrow \sqrt{\text{Var}[X^\tau]}$ as $n \rightarrow \infty$, and $\text{Var}[X^\tau] = \text E\big[ ( X^\tau - \text E[X^\tau] )^2\big]$. In practice the standard deviation, $\sigma^\tau_{n}$, is calculated using Welford's algorithm \cite{Welford1962}.
Since the MP2 energy is the sum over the independent expectation values of all $\tau$-points,
\begin{equation}
E_{\text x}^{(2)} = \frac 1 2 \int_0^\infty \text E [X^{\tau} ] \, \D\tau \approx \frac 1 2 \sum_\tau w_\tau \, \text E [X^{\tau} ] \;,
\end{equation}
the statistical error of the MP2 energy is simply estimated by the formula for the propagation of error,
\begin{equation}
\delta E_{\text x}^{(2)} = \frac 1 2 \sqrt{\sum_\tau (w_\tau \, \delta\mu^\tau_n)^2 } \;.
\end{equation}

In general, the system size scaling of the sample variance, $(\sigma^\tau_{n})^2$, plays an important role for the prefactor and the system size scaling of the computation time. If the variance obeys a polynomial system size scaling with the power $c_\text{var}$, we can conclude that the number, $n$, of necessary stochastic energies follows exactly the same scaling behavior if the statistical error should be kept constant [see Eq. (\ref{eq:stat_error})]. Moreover, let the calculation time of a single stochastic energy, $X^\tau$, have a polynomial system size scaling to the power of $c_\text{rnd}$. The total scaling of the computation time is then polynomial with the power $c_\text{var} + c_\text{rnd}$. Thus, due to Eq. (\ref{eq:stat_error}), the sample variance has a strong impact on both the scaling and the prefactor of the algorithm, making the variance the key quantity that determines whether the stochastic approach is competitive.

\subsection{Correlated sampling}\label{sec:correlsampling}

For correlated sampling we calculate a set of stochastic orbitals $|\kappa^\tau_\theta\rangle, |\lambda^\tau_\theta\rangle, |\alpha^\tau_\theta\rangle, |\beta^\tau_\theta\rangle$, as indicated by the new index $\theta=1,...,N_\theta$. Clearly, the vectors of random coefficients, $p_i, q_i, r_a, s_a$, in Eq. (\ref{eq:StochOrbit}) have to be replaced by matrices of random coefficients $p_{i\theta},q_{i\theta},r_{a\theta},s_{a\theta}$. To calculate a sample we can now write
\begin{equation}
X^\tau  =   \langle \kappa^\tau_\theta \lambda^\tau_{\theta'} | \alpha^\tau_\theta  \beta^\tau_{\theta'} \rangle \langle \alpha^\tau_\theta \beta^\tau_{\theta'}| \lambda^\tau_{\theta'} \kappa^\tau_{\theta}\rangle \;,  \label{eq:correlsample}
\end{equation}
which is equal to uncorrelated sampling of Eq. (\ref{eq:sample}) as long as $\theta=\theta'$, but activates correlated sampling when combinations of $\theta\neq\theta'$ are allowed. Hence, a larger sample can be calculated with the same amount of stochastic orbitals. \revcolI{Comparable sampling techniques for MP2 were applied in Ref. \cite{Willow2013,Ge2014}}. Since the generation of stochastic orbitals has a steeper system size scaling than the evaluation of the two-electron integrals, correlated sampling leads to a significant speed up. This will be discussed in detail in the next section.

%%%%%%%%%%%%%%%%%%%%%%%%%%%%%%%%%%%%%%%%%%%%%%%%%%%%%%%%%%%%%%%%%
\section{Implementation} \label{sec:implementation}
%%%%%%%%%%%%%%%%%%%%%%%%%%%%%%%%%%%%%%%%%%%%%%%%%%%%%%%%%%%%%%%%%

The algorithm presented in this paper is implemented in the Vienna ab initio simulation package (VASP). \revcolI{Since VASP is designed for periodic systems, it naturally rests upon the plane wave basis. The basis set size can thus easily be controlled by the plane wave cutoff $E_{\text{cut}}$. Hence, we can focus on the statistical fluctuations and no additional error has to be considered, as it appears in local correlation methods. On the other hand, we forgo the benefits of localized basis sets, which reportedly allow to sample energy differences accurately even using a small number of stochastic orbitals \cite{Neuhauser2017}.} Furthermore, as every method in VASP, the implementation is based on the projector augmented wave (PAW) method. However, for brevity, the PAW is ignored in all formulas of this work. 

\subsection{Algorithm and scaling} \label{sec:Algo+Scale}

\begin{figure}
\includegraphics{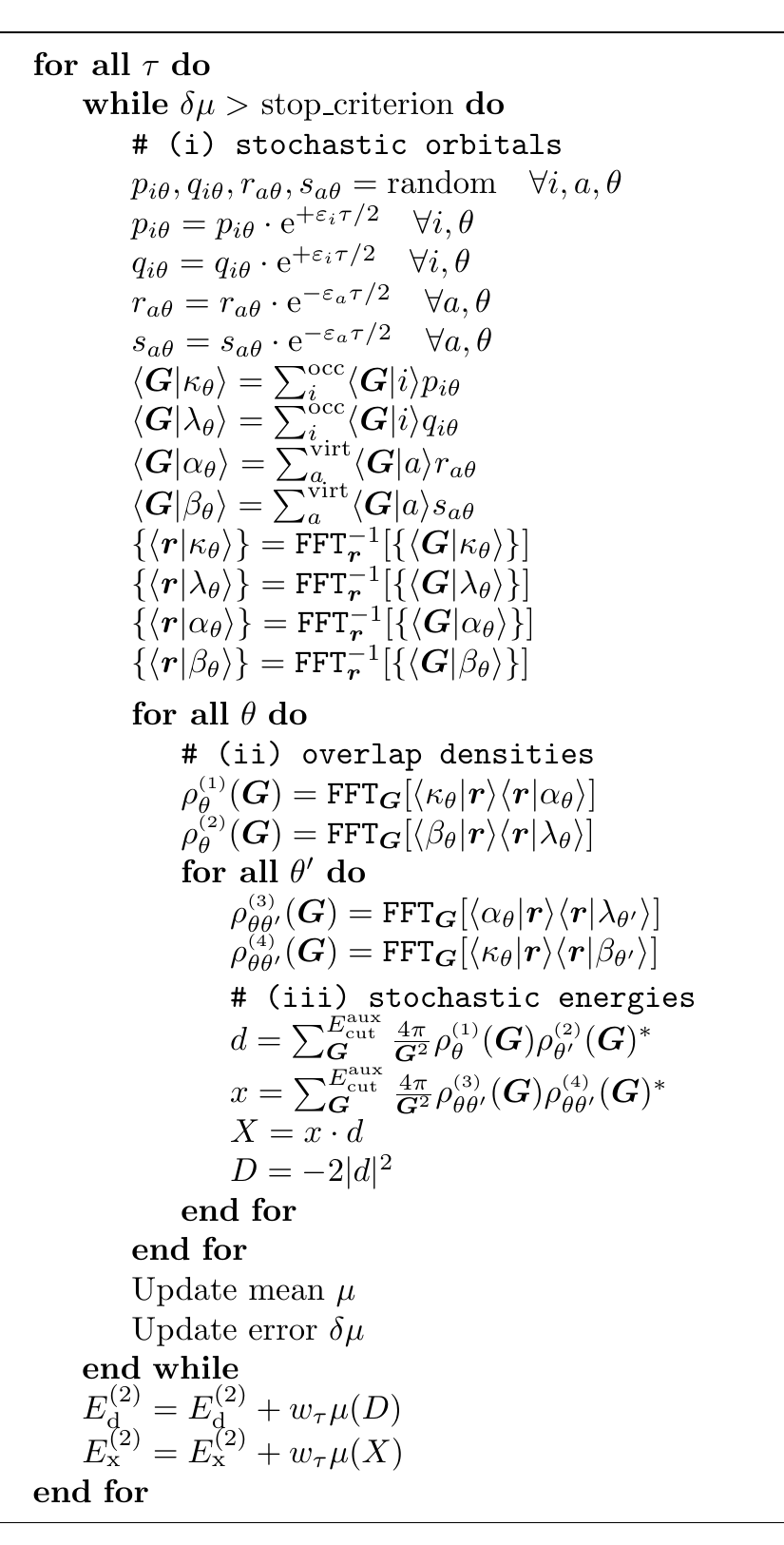}
\caption{Pseudocode of the MP2 algorithm with stochastic orbitals. Note that the HF orbitals, $|i\rangle , |a\rangle$, and energies, $\varepsilon_i, \varepsilon_a$, stem from a preceding HF calculation.}
\label{fig:pseudocode}
\end{figure}

The algorithm can be divided into three simple parts: for each $\tau$-point (i) calculate the stochastic orbitals (\ref{eq:StochOrbitPW}) from the HF orbitals, (ii) calculate the overlap densities (\ref{eq:overlap}), (iii) calculate the two-electron integrals (\ref{eq:SCH2EI}) using the overlap densities, and the stochastic energy (\ref{eq:sample}) and update the statistics. Repeat this procedure until the error of the mean for this $\tau$-point is below the desired threshold. In the following, correlated sampling is considered, i.e. the additional index $\theta$ comes into play, as introduced in Sec. \ref{sec:correlsampling}. Pseudocode of the algorithm can be found in Fig. \ref{fig:pseudocode}.

In step (i), $N_\theta$ stochastic orbitals are calculated using BLAS level 3 routines and stored in memory, where $N_\theta$ is user given. The scaling with computation time and memory reads
\begin{align}
|\kappa^\tau_\theta\rangle , |\lambda^\tau_\theta\rangle &\sim 
\begin{cases}
  \mathcal O(N_\theta N_{\bm G} N_i ) & \text{in time}  \\
  \mathcal O(N_\theta N_{\bm r} ) & \text{in memory,}  \label{eq:rnd_orbitals_scale}\\
\end{cases} \\
|\alpha^\tau_\theta\rangle , |\beta^\tau_\theta\rangle & \sim
\begin{cases}
  \mathcal O(N_\theta N_{\bm G} N_a ) & \text{in time}  \\
  \mathcal O(N_\theta N_{\bm r} ) & \text{in memory.}  \\
\end{cases} 
\end{align}
Here, $N_{\bm G}$ is the number of reciprocal lattice vectors limited by the cutoff $E_{\text{cut}}$, and $N_i$ and $N_a$ are the number of occupied and virtual HF orbitals, respectively. The number of real space grid points, $N_{\bm r}$, determines the memory scaling when the stochastic orbitals are Fourier transformed to real space to calculate the overlap densities. Since the number of $\tau$-points is largely independent of the system size, we ignore this factor in the analysis.

For step (ii), the following overlap densities have to be calculated for the stochastic energies (\ref{eq:correlsample}):
%In order to calculate all necessary overlap densities for step (ii), we exploit the fact, that with $\bm\Gamma$-only sampling of the Brillouine zone, the HF orbitals can be considered als real valued functions. Hence we can write $ \langle \alpha^\tau_\theta \beta^\tau_{\theta'}| \lambda^\tau_{\theta'} \kappa^\tau_{\theta}\rangle = \langle \lambda^\tau_{\theta'} \kappa^\tau_\theta | \alpha^\tau_\theta  \beta^\tau_{\theta'} \rangle$. Furthermore, this argument also leads to $\langle \lambda^\tau_\theta | \EuE^{+\ImI\bm G \hat {\bm r}} | \beta^\tau_\theta \rangle = \langle \lambda^\tau_\theta | \EuE^{-\ImI\bm G \hat {\bm r}} | \beta^\tau_\theta \rangle^{*}$. Thus, the following overlap densities are sufficient for the samples (\ref{eq:correlsample}):
\begin{align}
{\langle \kappa^\tau_\theta | \EuE^{-\ImI\bm G \hat {\bm r}} | \alpha^\tau_\theta \rangle} \atop {\langle \beta^\tau_\theta | \EuE^{-\ImI\bm G \hat {\bm r}} | \lambda^\tau_\theta \rangle} 
&\sim
\mathcal O(N_\theta N_{\bm G}^{\text{aux}} \ln N_{\bm G}^{\text{aux}} )  \quad \text{in time,}  \label{eq:uncorreloverlapscale}
\\[5pt]
{\langle \alpha^\tau_\theta | \EuE^{-\ImI\bm G \hat {\bm r}} | \lambda^\tau_{\theta'} \rangle} \atop {\langle \kappa^\tau_{\theta} | \EuE^{-\ImI\bm G \hat {\bm r}} | \beta^\tau_{\theta'} \rangle}  
&\sim 
\mathcal O(N_\theta^2 N_{\bm G}^{\text{aux}} \ln N_{\bm G}^{\text{aux}} )  \quad \text{in time,}  \label{eq:correloverlapscale}
\end{align}
where $N_{\bm G}^{\text{aux}}$ is the number of reciprocal lattice vectors limited by the auxiliary cutoff $E_\text{cut}^\text{aux}$. Note that the first two overlap densities involve only one $\theta$ index, whereas the last two overlap densities have to be calculated for combinations of $\theta$ and $\theta'$. It is also worth mentioning, that for the direct MP2 energy, only the first two overlap densities (\ref{eq:uncorreloverlapscale}) are necessary. Thus, in each loop cycle the algorithm checks, whether the accuracy of the exchange energy was already reached, in order to decide whether the computation of the more expensive overlap densities (\ref{eq:correloverlapscale}) is necessary. Since the variance of the direct MP2 energy turns out to be larger than the variance of the exchange energy, this approach leads to a significant speed up of the algorithm.

Having the overlap densities we can calculate the stochastic energies (iii) that scale as
\begin{equation}
\langle \kappa^\tau_\theta \lambda^\tau_{\theta'} | \alpha^\tau_\theta  \beta^\tau_{\theta'} \rangle \langle \alpha^\tau_\theta \beta^\tau_{\theta'}| \lambda^\tau_{\theta'} \kappa^\tau_{\theta}\rangle \sim \mathcal O(N_\theta^2 N_{\bm G}^{\text{aux}})  \label{eq:apply_gfac_scale}
\end{equation}
in time. The prefactor of this step (iii) is small compared to that of (ii), since step (ii) involves FFTs, whereas here only summations over reciprocal lattice vectors are performed.

With this at hand one can calculate the actual scaling of the algorithm with the system size (independent of $N_\theta$) and demonstrate the benefit of the correlated sampling. For a given system and fixed absolute statistical error, the sample size reads $n=n_\text{L}N_\theta^2$, where $n_\text{L}$ is the number of loop cycles of the steps (i)-(iii), since the variance is independent of $N_\theta$ (as will be shown in Sec. \ref{sec:correl_smpl_bench}). With (\ref{eq:rnd_orbitals_scale})-(\ref{eq:apply_gfac_scale}) we can then write the computation time as a function of $N_\theta$:
\begin{align}
T(N_\theta) &= 2\gamma_{\text{BLAS}} \frac{n}{N_\theta} N_{\bm G} (N_a+N_i)  && \text{(i)} \nonumber\\
           &+  2\gamma_{\texttt{FFT}}n \Big(1+\frac 1 N_{\theta}\Big)N_{\bm G}^{\text{aux}}\ln N_{\bm G}^{\text{aux}} && \text{(ii)} \label{eq:comptimecorrel}\\
           &+ \gamma nN_{\bm G}^{\text{aux}} && \text{(iii)} \nonumber \;.
\end{align}
Here $\gamma_{\text{BLAS}}$, $\gamma_{\texttt{FFT}}$, and $\gamma$ are the prefactors of the BLAS routines, fast Fourier transforms, and simple multiplications, respectively. Equation (\ref{eq:comptimecorrel}) shows clearly that correlated sampling (increasing $N_\theta$) asymptotically reduces the computation time. In this approach the largest possible $N_\theta$ is given by $n_\text{L}=1 \Rightarrow  n=N_\theta^2$ such that the entire sample is generated in one single loop cycle and only $N_\theta$ stochastic orbitals have to be generated for the MP2 calculation. In this optimal case the computation time reduces to 
\begin{align}
T(N_\theta=\sqrt{n}) &\approx 2\gamma_{\text{BLAS}} \sqrt{n} N_{\bm G} N_a  && \text{(i)} \nonumber\\
           &+  2\gamma_{\texttt{FFT}}nN_{\bm G}^{\text{aux}}\ln N_{\bm G}^{\text{aux}} && \text{(ii)} \label{eq:comptimecorrel_opt}\\
           &+ \gamma nN_{\bm G}^{\text{aux}} && \text{(iii)} \;, \nonumber
\end{align}
where we have assumed $N_i/N_a \ll 1$ and $1/\sqrt{n} \ll 1$. Later, in Sec. \ref{sec:varscale}, we will show that the sample variance and therefore also the sample size, $n$, scales quadratically with the system size, if a fixed absolute statistical error is imposed. Thus, Eq. (\ref{eq:comptimecorrel_opt}) shows, that each step of the algorithm possesses a cubic scaling with the system size.\\
If only a fixed relative error is required, then the sample size can be chosen independently of the system size and the scaling reduces to a largely quadratic scaling of step (i). In practice, however, the computation time is dominated by the linear scaling of step (ii), as long as
\begin{equation}
\frac{\gamma_{\texttt{FFT}}}{\gamma_{\text{BLAS}}} > 1.84 \cdot \frac{N_a}{\sqrt{n}}
\end{equation}
where $N_{\bm G} / N_{\bm G}^{\text{aux}}\approx 1.84$ was assumed, following from the mentioned ratio $E_\text{cut}^{\text{aux}}=2 E_\text{cut}/3$. It is only this scenario of a fixed relative error in combination with a moderate system size where approximate "linear scaling" can be expected.\\
Furthermore, we want to stress how the correlated sampling is responsible for the reduction to a cubical scaling in the case of a fixed absolute statistical error. For this, we look at the speed up which can be obtained by the correlated sampling,
\begin{equation}
\frac{T(N_\theta=1)}{T(N_\theta = \sqrt{n})} \sim N_a\;.
\end{equation}
We conclude, that the possible speed up due to correlated sampling increases linearly with the system size. This, and the fact that the variance and, therefore, the sample size scale quadratically, is the reason why a cubic scaling of the stochastic MP2 algorithm can be achieved. Without correlated sampling the generation of stochastic orbitals (i) would be the dominant step, yielding for the entire algorithm a quartic scaling behavior.

Also, the key differences between this approach and the method by Neuhauser \cite{Neuhauser2013} can now be summarized as follows. We use the exact HF orbitals in the plane wave basis as the starting point, instead of generating completely random functions that are purified to the occupied or unoccupied space. Our HF orbitals stem from a full HF calculation where the Fock-matrix is not to be assumed as sparse. The stochastic orbitals are propagated in imaginary time instead of real time by a simple multiplication and no application of the Fock matrix is necessary. Additionally, we introduced correlated sampling, which reduces the seemingly most expensive step of generating stochastic orbitals from HF orbitals to a less relevant contribution to the computation time (see Eq. (\ref{eq:comptimecorrel}) or Fig. (\ref{fig:correl_time})).

\subsection{Complex vs. real random coefficients} \label{sec:real_vs_complex}

In Sec. \ref{sec:LTMP2_with_rnd_orbtials} the stochastic orbitals were introduced as linear combinations of time-dependent HF orbitals with uniformly distributed random complex numbers as coefficients. Instead of complex random coefficients, also real random coefficients can be used. In the case of $\bm \Gamma$-only sampling of the Brillouine zone, the spatial HF orbitals are real, hence, the stochastic orbitals would inherit this property, if only real random coefficients are used. For calculations in the plane wave basis, real spatial orbitals are beneficial, since the overlap densities have to be calculated only for half of the number of plane wave vectors. This is a consequence of the identity $\langle \psi | \EuE^{-\ImI\bm G \hat{\bm r}}|\varphi\rangle = \langle \psi | \EuE^{+\ImI\bm G \hat{\bm r}}|\varphi\rangle^{*}$, for real spatial orbitals $\psi,\varphi$. Also, the orbitals are stored only for half of the plane wave coefficients, since $\langle \bm G | \varphi \rangle = \langle -\bm G | \varphi \rangle^{*} $ for any real spatial orbital $\varphi$. Hence, the computational effort to calculate a stochastic energy, $X^\tau$, is halved in time and memory for all three steps, (i)-(iii), of the algorithm. 

However, this comes at the price of a larger variance. This can be estimated by calculating the variance of the stochastic approximation of the Kronecker deltas, where the random coefficients were introduced initially. As described in Sec. \ref{sec:LTMP2_with_rnd_orbtials}, the Kronecker deltas are approximated as expectation values $\delta_{ii}=1=\text E[z^*z]$, where $z=a+\ImI b$ is a random complex number whose real and imaginary part are uniformly distributed over the range $[-\sqrt{3/2},+\sqrt{3/2}]$. The variance can be calculated as $\text{Var}[z^*z] = \text E[(z^*z - 1)^2] = 0.4$. If instead only real random coefficients are used, the Kronecker deltas are approximated as $\delta_{ii}=1=\text E[r^2]$, where $r$ is a uniformly distributed real random number over the range $[-\sqrt{3},+\sqrt{3}]$. But in this case the variance is twice as large as in the case of complex numbers, $\text{Var}[r^2] = \text E[(r^2 - 1)^2] = 0.8$. 

Analytically, it is not possible to conclude whether the variance of the stochastic MP2 algorithm also doubles, when real instead of complex random coefficients are employed. However, in Sec. \ref{sec:measured_var_real_cmplx} we present a comparison of stochastic MP2 runs using real and complex random coefficients, confirming that the variance roughly doubles for real random coefficients. 
%an approximate doubling of the stochastic MP2 variance. This means that the computation time is roughly equal for both cases, however, the memory requirement is still favorable in the case of real random numbers.

\subsection{Internal cutoff extrapolation}

It is known, that the correlation energy for wave-function based methods like MP2 or RPA (random phase approximation) converges slowly with respect to the basis set size (plane wave cutoff). However, the exact asymptotic behavior for large plane wave cutoffs is also know and can be exploited for an internal cutoff extrapolation. In this work, the cutoff that controls the basis set size of the calculations of the two-electron integrals (\ref{eq:SCH2EI}) is $E_\text{cut}^\text{aux}$ and according to Ref. \cite{Harl2008,Shepherd2012} the asymptotic behavior reads,
\begin{equation}
E^{(2)}(E_\text{cut}^\text{aux}) - E^{(2)}(E_\text{cut}^\text{aux}=\infty) \sim (E^\text{aux}_\text{cut})^{-3/2} \;. \label{eq:CutDecay}
\end{equation}
The internal cutoff extrapolation calculates the two-electron integrals for the user given cutoff $E^\text{aux}_\text{cut}$ and also for a certain number (8 in this work) of smaller cutoffs on the fly. This set of MP2 energies can be extrapolated to infinity according to Eq. (\ref{eq:CutDecay}). A detailed description of this extrapolation scheme can be found in Sec. III.D in Ref. \cite{Schafer2016}, where it was implemented for our deterministic quartic scaling MP2 algorithm.

\subsection{Parallelization} \label{sec:parallel}

For the parallelization, we use a combination of MPI and OpenMP. Since stochastic orbitals and overlap densities can be generated independently on all MPI ranks, the parallelization of the algorithm is rather simple and efficient. However, the access to the shared memory of the CPU sockets via OpenMP is favorable. Having the entire set of HF orbitals (occupied+unoccupied) available at each MPI rank, allows to calculate the stochastic orbitals (\ref{eq:StochOrbitPW}) without MPI communication. For large cells and large basis sets this requirement can quickly exceed the memory per single CPU, making shared memory the obvious solution. Hence, each MPI rank runs through the algorithm depicted in Fig. \ref{fig:pseudocode} and calculates correlated stochastic energies independently, whereas the OpenMP parallelization works differently in each of the steps (i)-(iii): In step (i) the BLAS routines are parallelized over the OpenMP threads (OpenMP aware BLAS). In step (ii) the \texttt{FFTs} are parallelized over the OpenMP threads (OpenMP aware FFT). And in step (iii) the sums over plane waves $\bm G$ are parallelized over the OpenMP threads.

%When, instead, k-point sampling is considered, the time-reversal symmetrie allows for the  

%%%%%%%%%%%%%%%%%%%%%%%%%%%%%%%%%%%%%%%%%%%%%%%%%%%%%%%%%%%%%%%%%
\section{Benchmark calculations} \label{sec:benchmark}
%%%%%%%%%%%%%%%%%%%%%%%%%%%%%%%%%%%%%%%%%%%%%%%%%%%%%%%%%%%%%%%%%

To compare the results with those of our recent publication of an exact quartic scaling MP2 algorithm \cite{Schafer2016} we, again, chose lithium hydride (LiH) and methane in a chabazite crystal as benchmark systems to test the parallelization efficiency, the system size scaling, the scaling of the variance, and the competitivity of the stochastic approach. All computations were performed on Intel Xeon E5-2650 v2 2.8 GHz processors. The timings are measured in CPU hours which is the CPU time in hours multiplied by the number of employed CPUs. We use VASP for all calculations, where we restrict on $\bm \Gamma$-only sampling of the Brillouine zone, a spin-restricted setting, and 6 $\tau$-points for the quadrature of the Laplace transform (\ref{eq:LTMP2}). If not explicitly stated, real random coefficients are used for the stochastic orbitals.

\subsection{Measured scaling of the sample variance} \label{sec:varscale}

\begin{figure}
\includegraphics[width=1.0\linewidth]{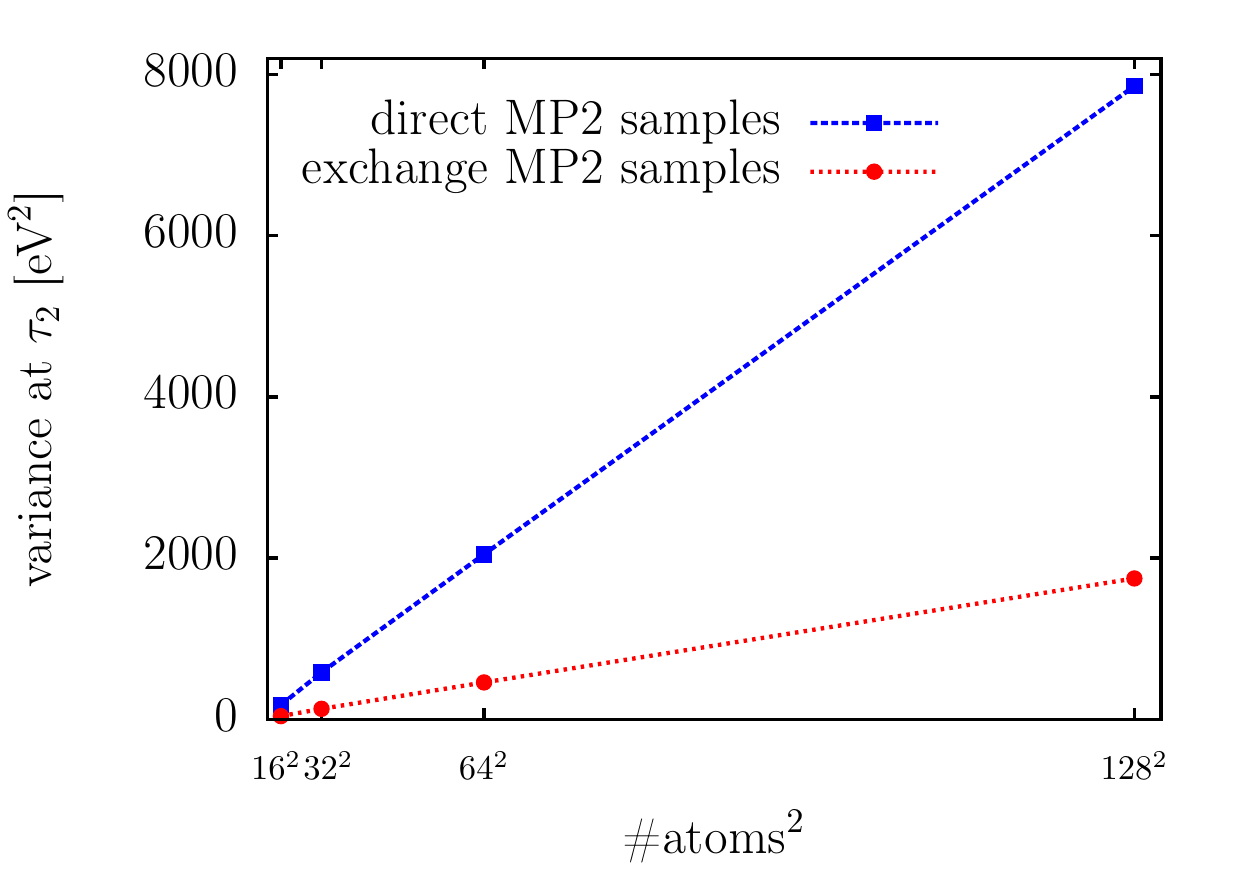}
\caption{Sample variance of the direct MP2 term at the second $\tau$-point for various supercells of LiH as a function of the number of atoms squared. This graph shows that the variance scales quadratically with the system size. }
\label{fig:var_sys_scaling}
\end{figure}

The variance is estimated by the sample variance, $(\sigma^\tau_{n})^2$, as described in Sec. \ref{sec:var_error}. In Fig. \ref{fig:var_sys_scaling} the sample variance at the second $\tau$-point of the $\tau$ quadrature (which gives the largest contribution to the MP2 energy) is plotted against the number of atoms squared. Since this results in a straight line, we can conclude that the variance possesses a quadratic scaling with the system size for each $\tau$ point. If the stopping criterion for the algorithm is given by a fixed absolute error, the sample size $n$ also needs to scale quadratically with the system size for each $\tau$ point, since the error behaves as $\sqrt{(\sigma^\tau_{n})^2/n}$.

\subsection{Correlated sampling} \label{sec:correl_smpl_bench}

\begin{figure}
\includegraphics[width=1.0\linewidth]{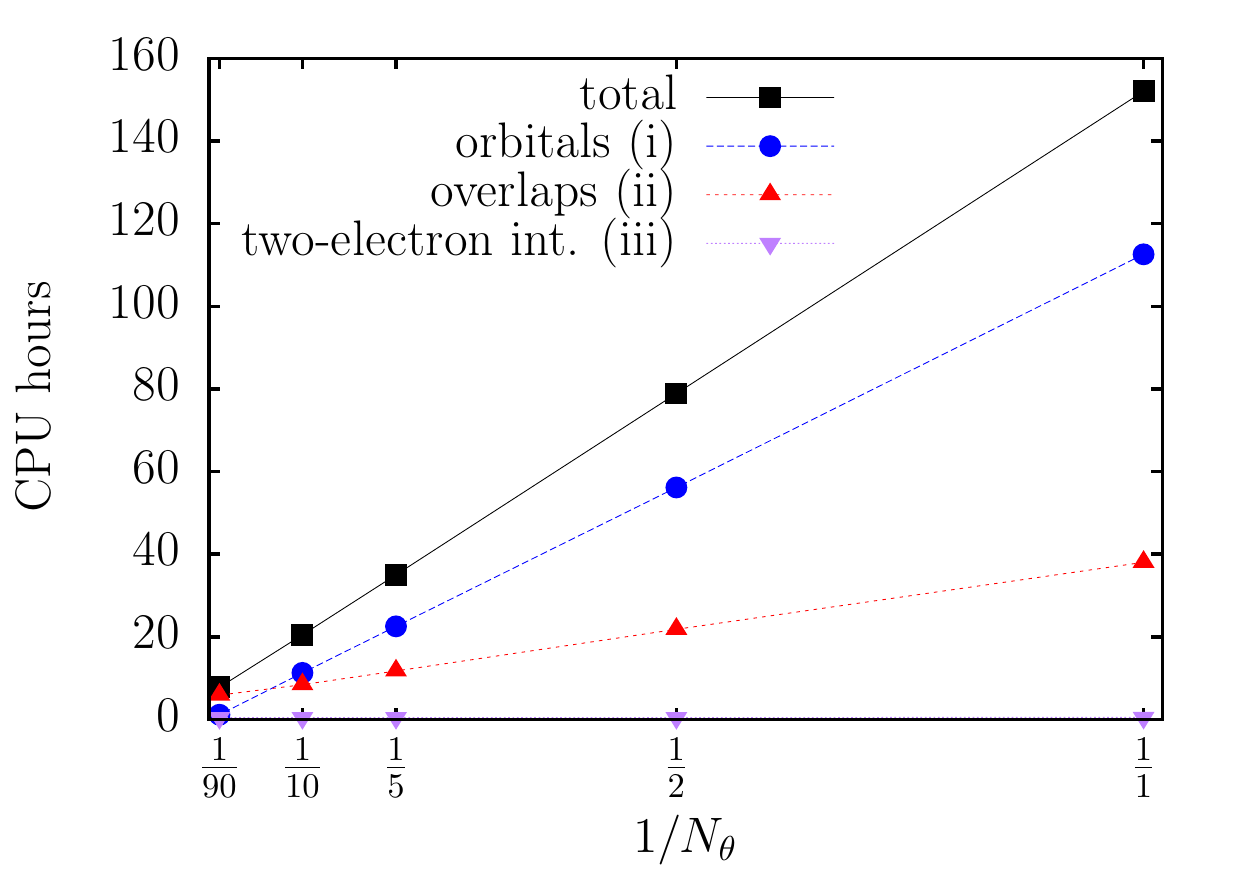}
\caption{Reducing the computation time using correlated sampling. The MP2 energy of a LiH supercell containing 32 atoms was calculated for different $N_\theta$, controlling the correlated sampling. The measured timings are in agreement with the prediction in Eq. (\ref{eq:comptimecorrel}).}
\label{fig:correl_time}
\end{figure}

In a previous section we derived that correlated sampling reduces the computation time to a plateau, see Eq. (\ref{eq:comptimecorrel}). We put this equation to test with a supercell of solid LiH containing 32 atoms. The computation time of each step of the algorithm, (i)-(iii) (see Sec. \ref{sec:Algo+Scale}), was measured against $N_\theta$, which controls the correlated sampling such that $N_\theta^2$ stochastic energies are calculated in each loop cycle. In Fig. \ref{fig:correl_time} the computation time is plotted against $1/N_\theta$, providing an apparent verification of the $1/N_\theta$ law of Eq. (\ref{eq:comptimecorrel}).

\begin{figure}
\includegraphics[width=1.0\linewidth]{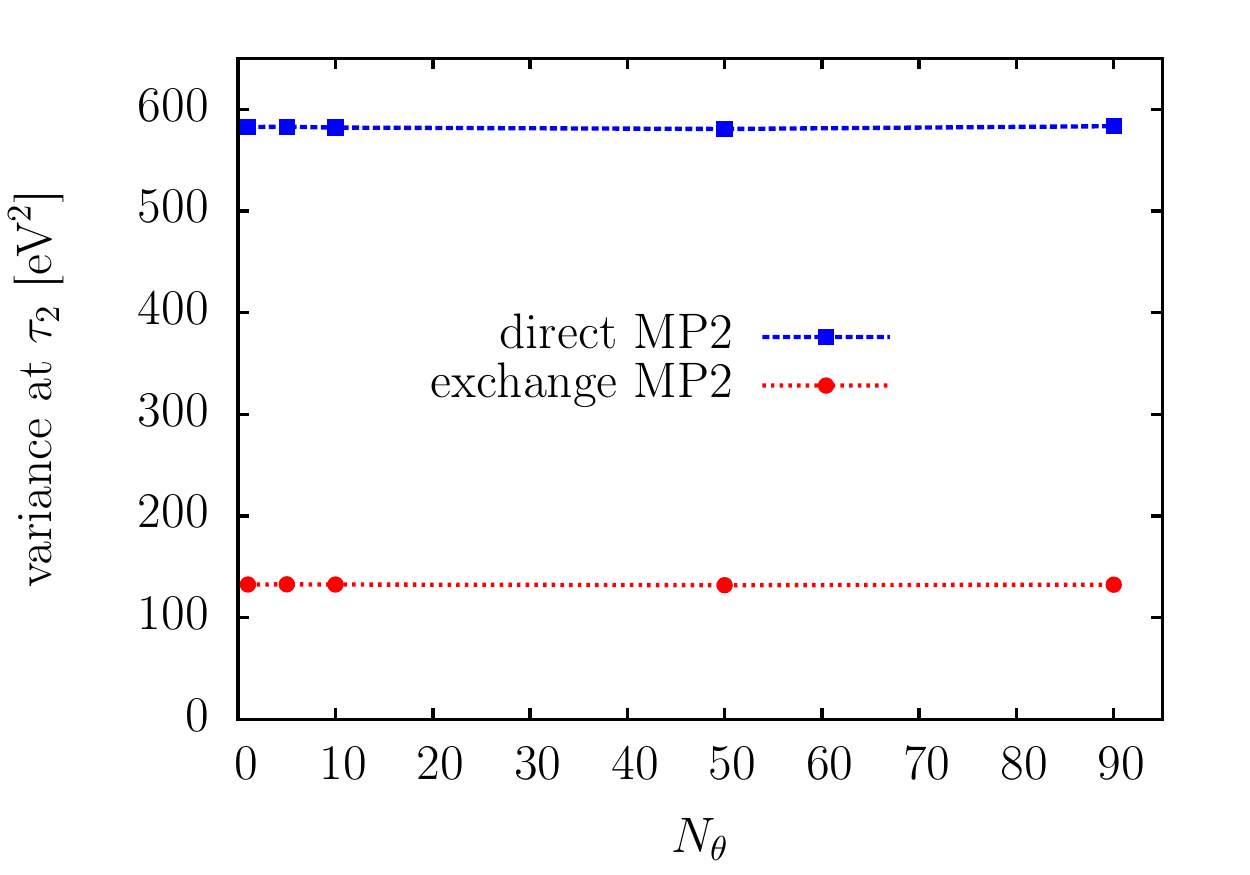}
\caption{Sample variance at the second $\tau$-point for different $N_\theta$. For both, the direct and exchange part, the variance is independent of the correlated sampling.}
\label{fig:correl_var}
\end{figure}

Furthermore, \emph{a priori} it is not clear if the correlated sampling affects the variance. To probe this, we plot the sample variance of the second $\tau$-point for both the direct and exchange MP2 energy against $N_\theta$, using the same benchmark system. The result can be seen in Fig. \ref{fig:correl_var}. Apparently, there is no visible effect of the correlated sampling, aside from stochastic fluctuations which are smaller than $0.7\,\%$ . Thus, we assume the variance to be constant with respect to $N_\theta$.

\subsection{Measured system size scaling and memory consumption} 

\begin{figure}
\includegraphics[width=1.0\linewidth]{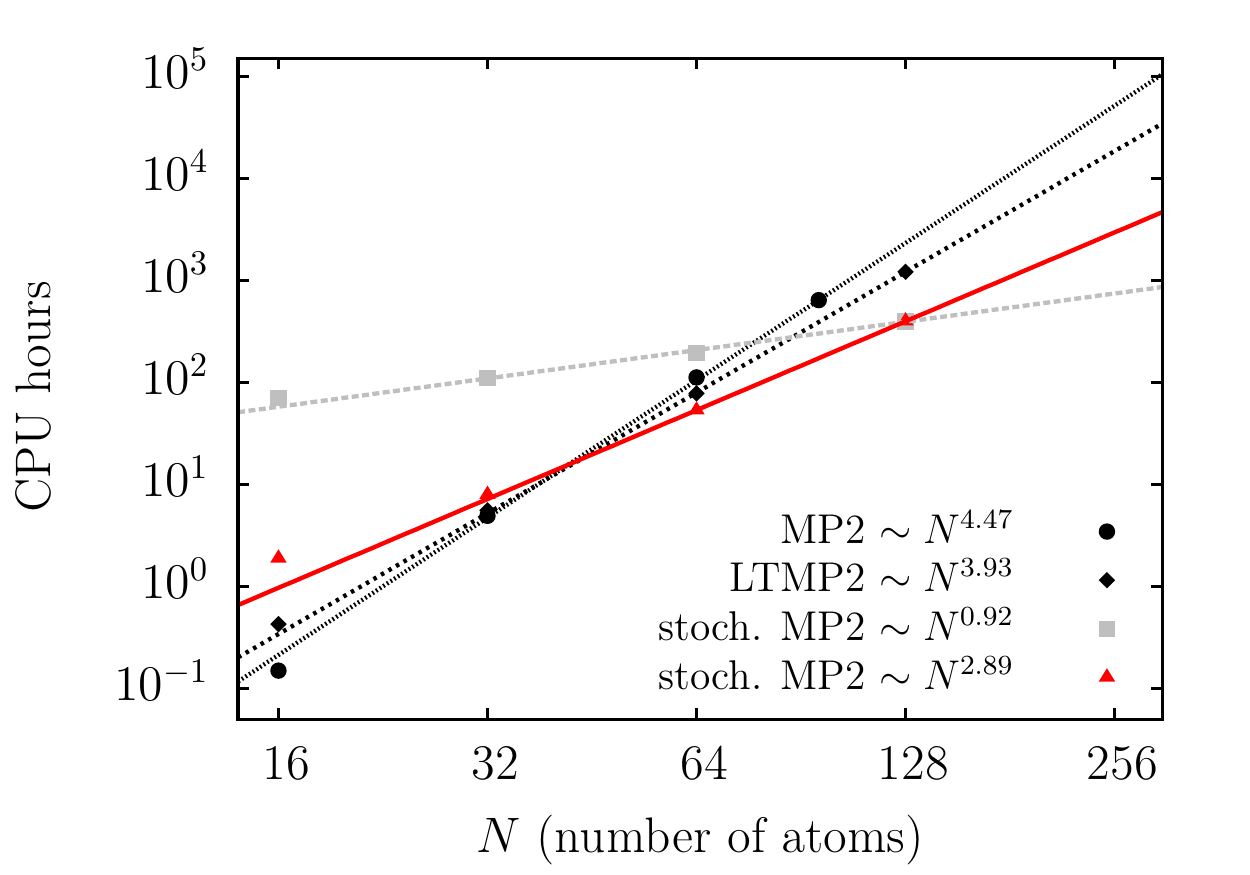}
\caption{System size scaling of different MP2 algorithms in VASP for various supercells of solid LiH. The red triangles are stochastic MP2 calculations with a fixed absolute error of $17.3\,\text{meV}$. The gray squares are stochastic MP2 calculations with a fixed relative error of $0.135 \, \text{meV per valence band}$. The black squares are calculations using the deterministic quintic scaling algorithm \cite{Marsman2009, Gruneis2010} and the black diamond symbols refer to the recently published deterministic quartic scaling MP2 algorithm (LTMP2) \cite{Schafer2016}. }
\label{fig:sys_scale}
\end{figure}

In order to measure the scaling of the stochastic MP2 algorithm with the system size, calculations on various supercells of LiH were performed. Figure \ref{fig:sys_scale} shows the result. As predicted in section \ref{sec:Algo+Scale}, the stochastic MP2 approach possesses a roughly cubic scaling, if a fixed absolute error is imposed. For a fixed relative error (per valence orbital) the scaling is roughly linear in the measured range of atoms. Detailed computational settings are provided in Tab. \ref{tab:memory_cons}. We also checked the sensitivity of the variance if the symmetry of a cell is broken. Therefore, we performed the calculation with 32 LiH atoms also with a supercell containing slightly displaced atoms, corresponding to a snapshot of a supercell at $1000\,\text K$. For each $\tau$ point we found no measurable effect on the variance, besides fluctuations around 1\%. 

If the statistical error should be decreased by a factor $c$, then the computation time increases by a factor of $c^2$, since the the statistical error decreases as $1/\sqrt {n}$, as was mentioned in the last section. Thus the break point of the stochastic approach compared to the deterministic codes depends mostly on the desired accuracy. Increasing, e.g., the accuracy of the stochastic calculations in Fig. \ref{fig:sys_scale} by a factor of 10 would shift the stochastic lines upwards by a factor of 100 in the computation time.

Regarding memory consumption the major contribution stems from the HF orbitals and the stochastic orbitals. In Tab. \ref{tab:memory_cons} the computational settings and the memory consumption is presented for three systems, including methane in a chabazite crystal and two different supercells of LiH. Of course, the memory requirements could be lowered considerably if the HF orbitals are distributed over all MPI ranks instead of only over the OpenMP threads, however, then the calculation of the stochastic orbitals (i) would require MPI communication, which would lower the parallelization efficiency.

\begin{table}
\begin{tabular*}{\linewidth}{l@{\extracolsep{\fill}}rrr}
\hline\hline
                                  & CH${}_4$ in Chab. &       LiH       &    LiH       \\
\hline
\#atoms                           &     $42$          &       $32$       &    $128$     \\
$E_\text{cut}$                    &    $750$ eV       &     $434$ eV     &   $434$ eV   \\
$E_\text{cut}^\text{aux}$         &    $500$ eV       &     $289$ eV     &   $289$ eV   \\
$N_{\bm G}$                       &    $18\,873$      &     $2\,563$     &   $10\,263$  \\
$N_{\bm G}^\text{aux}$            &    $10\,231$      &     $1\,418$     &   $5\,559$   \\
$N_{\bm r}$                       &    $74\,088$      &     $10\,976$    &   $43\,904$           \\
$N_i$                             &    $100$          &     $32$         &   $128$      \\
$N_a$                             &    $37\,660$      &     $5\,104$     &   $20\,400$  \\
$N_\theta$                        &   $150$           &     $120$        &   $240$      \\
$n$ for $E^{(2)}_d$         & $5.3 \cdot 10^9$  & $5.9 \cdot 10^7$ & $7.4 \cdot 10^8$       \\ 
$n$ for $E^{(2)}_x$         & $0.9 \cdot 10^9$  & $1.5 \cdot 10^7$ & $1.7 \cdot 10^8$       \\ 
standard deviation                &    $9.2$ meV      &     $13.8$ meV   &   $15.0$ meV \\
CPU hours                         &    $7\,121$       &     $8$          &   $398$      \\
HF orbitals                       &   $11.5$ GB       &   $238.2$ MB     &   $3.8$ GB   \\
stoch. orbitals                   &   $1.1$ GB        &   $130.3$ MB     &   $1.1$ GB   \\
total memory                      &   $13.4$ GB       &   $692.0$ MB     &   $6.1$ GB  \\ 
\hline\hline
\end{tabular*}
\caption{Computational settings and costs for different benchmark systems. The memory consumption is listed per MPI rank. The difference between the total memory and the orbital memory consumption is also due to the module for the statistic that temporarily stores the stochastic energies until the next MPI communication. The standard deviation refers to the statistical standard deviation of the MP2 energy, hence, all calculation are converged well below 1 meV per valence orbital, where $N_i$ is the number of occupied/valence orbitals. The sample size is denoted by $n$.} 
\label{tab:memory_cons}
\end{table}

\subsection{Measured parallelization efficiency}

\begin{figure}
\includegraphics[width=1.0\linewidth]{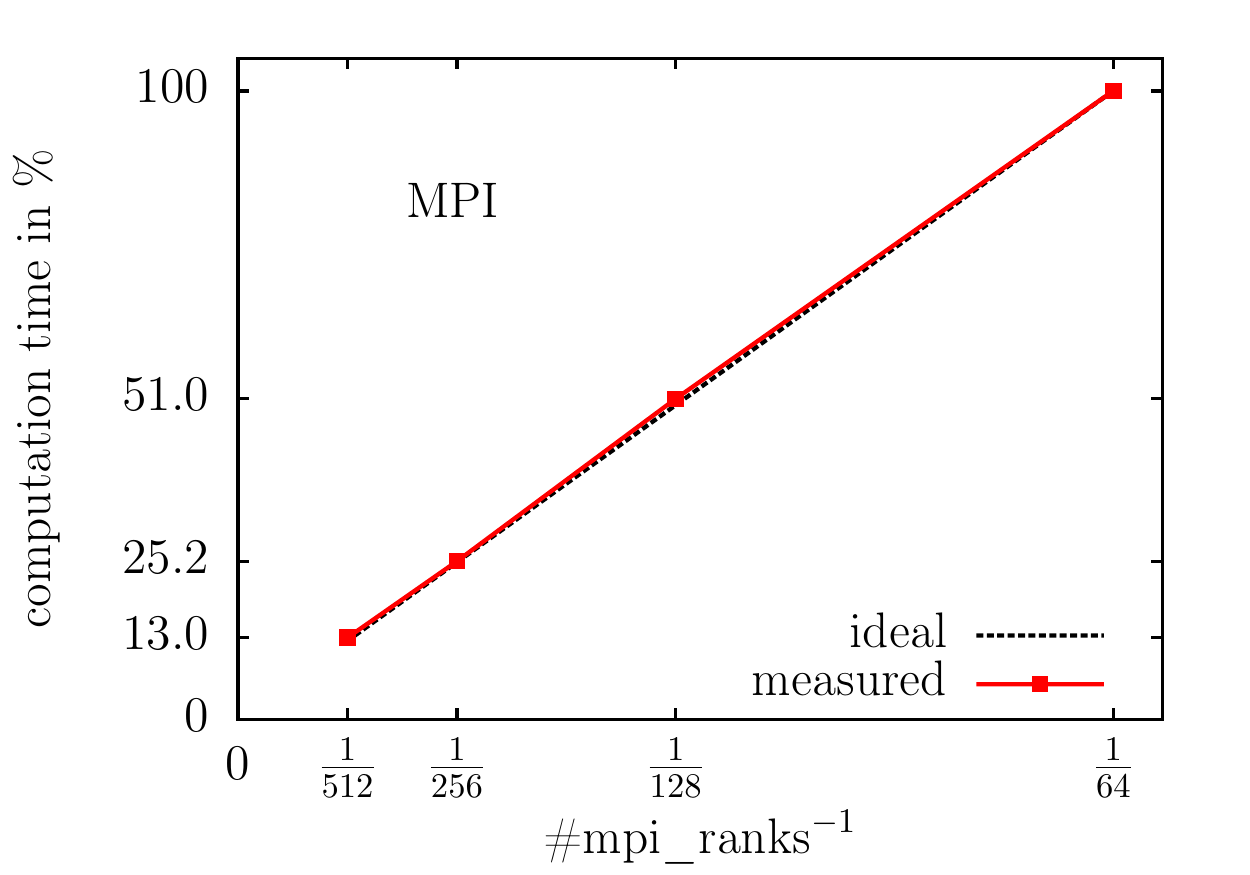}\\
\includegraphics[width=1.0\linewidth]{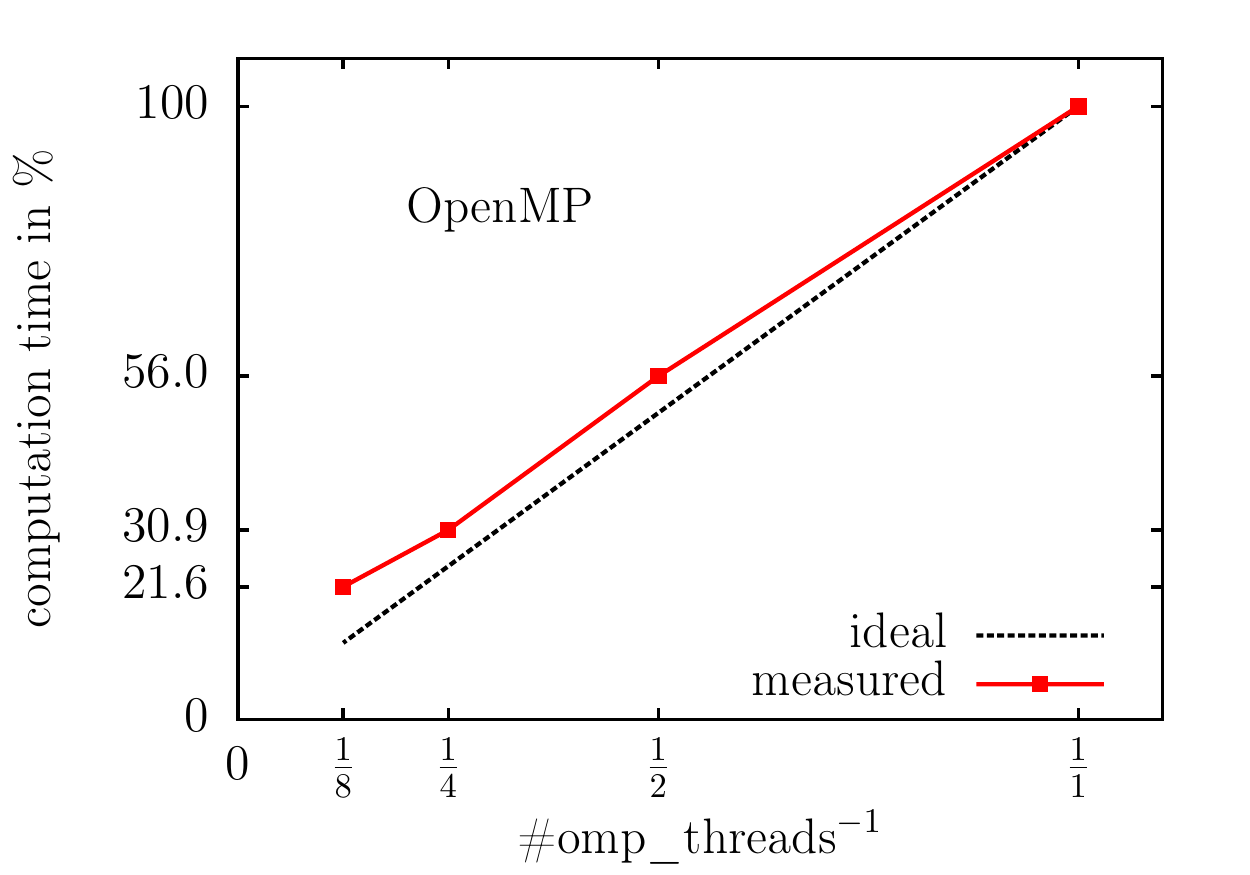}
\caption{Strong scaling of the MPI (upper graph) and the OpenMP parallelization (lower graph) for a stochastic MP2 calculation of a supercell of solid LiH containing 128 (upper) and 32 (lower) atoms. In all calculations a fixed number of stochastic energies was generated for each $\tau$-point. In the upper graph the number of OpenMP threads was fixed to 1 whereas in the lower graph the number of MPI ranks was fixed to 16.}
\label{fig:parallel_scaling}
\end{figure}

The parallelization with MPI and OpenMP is implemented as described in Sec. \ref{sec:parallel}. We measured the strong scaling of the stochastic MP2 algorithm for both MPI and OpenMP separately. As benchmark systems we used supercells of solid LiH with 128 atoms for the MPI scaling and with 32 atoms for the OpenMP scaling. Figure \ref{fig:parallel_scaling} shows the result. As expected, the parallelization with MPI is almost ideal, since communication is necessary only when the statistics are updated and all stochastic energies have to be gathered from all MPI ranks. The OpenMP parallelization, which is useful only if shared memory is required, shows also a tolerable strong scaling. The lower efficiency is a consequence of the simultaneous memory access of the OpenMP threads as well as the OpenMP overhead in e.g. OpenMP aware BLAS \revcolI{and FFT routines}.

\subsection{Measured variance with real and complex random coefficients} \label{sec:measured_var_real_cmplx}

How real numbers instead of complex random coefficients affect the variance, as discussed in Sec. \ref{sec:real_vs_complex}, was measured for four different systems. Since the presented algorithm calculates the variance for each $\tau$-point individually, we compare the variance of the exchange MP2 energy again at the second $\tau$-point, $\tau_2$. Table \ref{tab:real_cplx_var} shows that, for large systems, the variance roughly doubles, as supposed, when changing from complex to real random coefficients. However, for smaller systems, as for the methane molecule, the variance increases by a factor of $3.1$ for real coefficients, making the calculation slower by a factor of $\sqrt{3.1/2}=1.24$ at this $\tau$-point. It seems, that complex coefficients are beneficial for small systems, where the larger memory consumption is unproblematic and the speed up can be exploited. For larger systems, real random coefficients outperform the approach with complex random coefficients, since the effect on the computation time is negligible but the memory consumption is halved.

\begin{table}
\begin{tabular*}{\linewidth}{l@{\extracolsep{\fill}}rrr}
\hline\hline
 & \multicolumn{2}{c}{Variance at $\tau_2$ [$\text{eV}^2$]} &\\
System & real  & complex & ratio  \\
\hline 
CH${}_4$ in Chab. & $3664.36$ & $1736.57$ & $2.1$ \\
CH${}_4$          & $9.38$    & $3.03$    & $3.1$ \\
LiH (32 atoms)   & $133.01$  & $56.21$ & $2.4$  \\
LiH (128 atoms)  & $1748.56$ & $829.18$ & $2.1$ \\
\hline\hline
\end{tabular*}
\caption{Ratio of the sample variance at the second $\tau$ point of the exchange MP2 part for calculations with real and complex coefficients for the stochastic orbitals.} 
\label{tab:real_cplx_var}
\end{table}

\subsection{Adsorption energy of methane in a chabazite crystal cage}

In our previous publication \cite{Schafer2016}, where we presented a deterministic and quartic scaling MP2 algorithm for solids (LTMP2), we calculated the MP2 correlation part of the adsorption energy of a methane molecule (CH${}_4$) in a chabazite crystal ($\text{Al}{\kern 0.1em}\text{H}{\kern 0.1em}\text{O}_{24}{\kern 0.1em}\text{Si}_{11} $) cage. The geometry of the chabazite crystal and the position of the methane molecule are taken from  \cite{Goltl2012} and then reoptimized using the optB88-vdW functional \cite{Klimes2010}. We repeated the calculation with the presented stochastic MP2 approach, using the exact same computational settings (see Tab. \ref{tab:memory_cons} for CH${}_4$ in Chab.). The calculation consists of three steps, calculating the MP2 correlation energy of the bare methane molecule, the bare chabazite cage, and the methane molecule inside the chabazite cage. Table \ref{tab:ch4_in_chab} shows the result for the adsorption energy and the computation time for the stochastic MP2 approach, and the two other MP2 algorithms \cite{Marsman2009,Schafer2016} in VASP. Those can be used as a reference. If an error below 5\% is required, the stochastic MP2 approach outperforms the deterministic quintic scaling algorithm \cite{Marsman2009} but is not competitive compared to the deterministic quartic scaling MP2 code \cite{Schafer2016}. The stochastic algorithm is favourable only if an error of about 20\% is accepted. 

%This demonstrates a major problem of stochastic approaches. Even though the scaling can be reduced, the prefactor depends strongly of the sample variance and limits the 

\begin{table}
\begin{tabular*}{\linewidth}{l@{\extracolsep{\fill}}ccr}
\hline\hline
algo.   & $E_\text{ad}^{(2)}$ &  $E_\text{ad}^{(2)}|_{E_\text{cut}^\text{aux}\rightarrow\infty}$ & CPU hours   \\
\hline
MP2 \cite{Marsman2009}   & $298.98$      &     n/a            &   $17\,448$    \\
LTMP2 \cite{Schafer2016} & $299.02$      & $294.83$         &  $3\,769$  \\
stoch. MP2               & $254 \pm 50$ & $261 \pm 50$ &  $788$ \\
stoch. MP2               & $318 \pm 23$ & $316 \pm 23$ &  $3\,518$ \\
stoch. MP2               & $287 \pm 15$ & $281 \pm 15$ &  $10\,988$ \\
%stoch. MP2               & $287 \pm 15$ & $281 \pm 65$ &  $10\,988$ \\
\hline\hline
\end{tabular*}
\caption{Correlation part of the adsorption energy in meV of a methane molecule in a chabazite crystal cage using three different MP2 algorithms in VASP. Here, $E_\text{ad}^{(2)}|_{E_\text{cut}^\text{aux}\rightarrow\infty}$ stands for the result of the internal cutoff extrapolation.} 
\label{tab:ch4_in_chab}
\end{table}

%%%%%%%%%%%%%%%%%%%%%%%%%%%%%%%%%%%%%%%%%%%%%%%%%%%%%%%%%%%%%%%%%
\section{Conclusion}
%%%%%%%%%%%%%%%%%%%%%%%%%%%%%%%%%%%%%%%%%%%%%%%%%%%%%%%%%%%%%%%%%

We implemented a stochastic algorithm to calculate the MP2 correlation energy of three dimensional periodic systems in VASP. The parallelization is highly efficient such that thousands of CPUs can be used. In principal, the exact MP2 energy can be reached employing sufficiently large samples for each $\tau$-point and the internal basis set extrapolation. We found a cubic scaling with the system size, if a fixed absolute statistical error is required. Linear scaling could be reached for a fixed relative error per valence orbital. This is a consequence of the quadratic scaling of the variance with the system size. We also demonstrated the limits of the stochastic approach by calculating the adsorption energy of a methane molecule in a chabazite crystal cage. If errors of about 20\% are acceptable for this calculation, the stochastic approach is preferable, however, an error of about 5\% already leads to a higher computational effort than for the deterministic quartic scaling algorithm. 

We believe that stochastic approaches are indeed a promising way to handle the high computational effort of MP2 calculations, however, we emphasize, that the advantage of the lower complexity can quickly be compensated by a larger prefactor when high accuracies are required. It is indispensable to develop techniques that considerably reduce the sample variance, if statistical errors below 1 meV per valance band are required.

\section{Acknowledgments}

Funding by the Austrian Science Fund (FWF) within the SFB ViCoM (F41) is grateful acknowledged. Computations were performed on the Vienna Scientific Cluster, VSC3.

\bibliography{MP2}

\begin{thebibliography}{52}%
\makeatletter
\providecommand \@ifxundefined [1]{%
 \@ifx{#1\undefined}
}%
\providecommand \@ifnum [1]{%
 \ifnum #1\expandafter \@firstoftwo
 \else \expandafter \@secondoftwo
 \fi
}%
\providecommand \@ifx [1]{%
 \ifx #1\expandafter \@firstoftwo
 \else \expandafter \@secondoftwo
 \fi
}%
\providecommand \natexlab [1]{#1}%
\providecommand \enquote  [1]{``#1''}%
\providecommand \bibnamefont  [1]{#1}%
\providecommand \bibfnamefont [1]{#1}%
\providecommand \citenamefont [1]{#1}%
\providecommand \href@noop [0]{\@secondoftwo}%
\providecommand \href [0]{\begingroup \@sanitize@url \@href}%
\providecommand \@href[1]{\@@startlink{#1}\@@href}%
\providecommand \@@href[1]{\endgroup#1\@@endlink}%
\providecommand \@sanitize@url [0]{\catcode `\\12\catcode `\$12\catcode
  `\&12\catcode `\#12\catcode `\^12\catcode `\_12\catcode `\%12\relax}%
\providecommand \@@startlink[1]{}%
\providecommand \@@endlink[0]{}%
\providecommand \url  [0]{\begingroup\@sanitize@url \@url }%
\providecommand \@url [1]{\endgroup\@href {#1}{\urlprefix }}%
\providecommand \urlprefix  [0]{URL }%
\providecommand \Eprint [0]{\href }%
\providecommand \doibase [0]{http://dx.doi.org/}%
\providecommand \selectlanguage [0]{\@gobble}%
\providecommand \bibinfo  [0]{\@secondoftwo}%
\providecommand \bibfield  [0]{\@secondoftwo}%
\providecommand \translation [1]{[#1]}%
\providecommand \BibitemOpen [0]{}%
\providecommand \bibitemStop [0]{}%
\providecommand \bibitemNoStop [0]{.\EOS\space}%
\providecommand \EOS [0]{\spacefactor3000\relax}%
\providecommand \BibitemShut  [1]{\csname bibitem#1\endcsname}%
\let\auto@bib@innerbib\@empty
%</preamble>
\bibitem [{\citenamefont {M{\o}ller}\ and\ \citenamefont
  {Plesset}(1934)}]{Moeller1934}%
  \BibitemOpen
  \bibfield  {author} {\bibinfo {author} {\bibfnamefont {C.}~\bibnamefont
  {M{\o}ller}}\ and\ \bibinfo {author} {\bibfnamefont {M.~S.}\ \bibnamefont
  {Plesset}},\ }\href {\doibase 10.1103/PhysRev.46.618} {\bibfield  {journal}
  {\bibinfo  {journal} {Phys. Rev.}\ }\textbf {\bibinfo {volume} {46}},\
  \bibinfo {pages} {618} (\bibinfo {year} {1934})}\BibitemShut {NoStop}%
\bibitem [{\citenamefont {Szabo}\ and\ \citenamefont
  {Ostlund}(1996)}]{Szabo1996}%
  \BibitemOpen
  \bibfield  {author} {\bibinfo {author} {\bibfnamefont {A.}~\bibnamefont
  {Szabo}}\ and\ \bibinfo {author} {\bibfnamefont {N.~S.}\ \bibnamefont
  {Ostlund}},\ }\href@noop {} {\emph {\bibinfo {title} {{Modern Quantum
  Chemistry}}}}\ (\bibinfo  {publisher} {Dover Publications, Inc},\ \bibinfo
  {year} {1996})\BibitemShut {NoStop}%
\bibitem [{\citenamefont {Willow}\ \emph {et~al.}(2012)\citenamefont {Willow},
  \citenamefont {Kim},\ and\ \citenamefont {Hirata}}]{Willow2012}%
  \BibitemOpen
  \bibfield  {author} {\bibinfo {author} {\bibfnamefont {S.~Y.}\ \bibnamefont
  {Willow}}, \bibinfo {author} {\bibfnamefont {K.~S.}\ \bibnamefont {Kim}}, \
  and\ \bibinfo {author} {\bibfnamefont {S.}~\bibnamefont {Hirata}},\ }\href
  {\doibase 10.1063/1.4768697} {\bibfield  {journal} {\bibinfo  {journal} {J.
  Chem. Phys.}\ }\textbf {\bibinfo {volume} {137}} (\bibinfo {year} {2012}),\
  10.1063/1.4768697}\BibitemShut {NoStop}%
\bibitem [{\citenamefont {Alml{\"{o}}f}(1991)}]{Almlof1991}%
  \BibitemOpen
  \bibfield  {author} {\bibinfo {author} {\bibfnamefont {J.}~\bibnamefont
  {Alml{\"{o}}f}},\ }\href {\doibase 10.1016/0009-2614(91)80078-C} {\bibfield
  {journal} {\bibinfo  {journal} {Chem. Phys. Lett.}\ }\textbf {\bibinfo
  {volume} {181}},\ \bibinfo {pages} {319} (\bibinfo {year}
  {1991})}\BibitemShut {NoStop}%
\bibitem [{\citenamefont {Sch{\"{a}}fer}\ \emph {et~al.}(2017)\citenamefont
  {Sch{\"{a}}fer}, \citenamefont {Ramberger},\ and\ \citenamefont
  {Kresse}}]{Schafer2016}%
  \BibitemOpen
  \bibfield  {author} {\bibinfo {author} {\bibfnamefont {T.}~\bibnamefont
  {Sch{\"{a}}fer}}, \bibinfo {author} {\bibfnamefont {B.}~\bibnamefont
  {Ramberger}}, \ and\ \bibinfo {author} {\bibfnamefont {G.}~\bibnamefont
  {Kresse}},\ }\href {\doibase 10.1063/1.4976937} {\bibfield  {journal}
  {\bibinfo  {journal} {J. Chem. Phys.}\ }\textbf {\bibinfo {volume} {146}},\
  \bibinfo {pages} {104101} (\bibinfo {year} {2017})}\BibitemShut {NoStop}%
\bibitem [{\citenamefont {Manby}\ \emph {et~al.}(2006)\citenamefont {Manby},
  \citenamefont {Alf{\`{e}}},\ and\ \citenamefont {Gillan}}]{Manby2006}%
  \BibitemOpen
  \bibfield  {author} {\bibinfo {author} {\bibfnamefont {F.~R.}\ \bibnamefont
  {Manby}}, \bibinfo {author} {\bibfnamefont {D.}~\bibnamefont {Alf{\`{e}}}}, \
  and\ \bibinfo {author} {\bibfnamefont {M.~J.}\ \bibnamefont {Gillan}},\
  }\href {\doibase 10.1039/b613676a} {\bibfield  {journal} {\bibinfo  {journal}
  {Phys. Chem. Chem. Phys.}\ }\textbf {\bibinfo {volume} {8}},\ \bibinfo
  {pages} {5178} (\bibinfo {year} {2006})}\BibitemShut {NoStop}%
\bibitem [{\citenamefont {Casassa}\ \emph {et~al.}(2008)\citenamefont
  {Casassa}, \citenamefont {Halo},\ and\ \citenamefont
  {Maschio}}]{Casassa2008}%
  \BibitemOpen
  \bibfield  {author} {\bibinfo {author} {\bibfnamefont {S.}~\bibnamefont
  {Casassa}}, \bibinfo {author} {\bibfnamefont {M.}~\bibnamefont {Halo}}, \
  and\ \bibinfo {author} {\bibfnamefont {L.}~\bibnamefont {Maschio}},\ }\href
  {\doibase 10.1088/1742-6596/117/1/012007} {\bibfield  {journal} {\bibinfo
  {journal} {J. Phys. Conf. Ser.}\ }\textbf {\bibinfo {volume} {117}},\
  \bibinfo {pages} {12007} (\bibinfo {year} {2008})}\BibitemShut {NoStop}%
\bibitem [{\citenamefont {Halo}\ \emph
  {et~al.}(2009{\natexlab{a}})\citenamefont {Halo}, \citenamefont {Casassa},
  \citenamefont {Maschio},\ and\ \citenamefont {Pisani}}]{Halo2009}%
  \BibitemOpen
  \bibfield  {author} {\bibinfo {author} {\bibfnamefont {M.}~\bibnamefont
  {Halo}}, \bibinfo {author} {\bibfnamefont {S.}~\bibnamefont {Casassa}},
  \bibinfo {author} {\bibfnamefont {L.}~\bibnamefont {Maschio}}, \ and\
  \bibinfo {author} {\bibfnamefont {C.}~\bibnamefont {Pisani}},\ }\href
  {\doibase 10.1016/j.cplett.2008.11.043} {\bibfield  {journal} {\bibinfo
  {journal} {Chem. Phys. Lett.}\ }\textbf {\bibinfo {volume} {467}},\ \bibinfo
  {pages} {294} (\bibinfo {year} {2009}{\natexlab{a}})}\BibitemShut {NoStop}%
\bibitem [{\citenamefont {Halo}\ \emph
  {et~al.}(2009{\natexlab{b}})\citenamefont {Halo}, \citenamefont {Casassa},
  \citenamefont {Maschio},\ and\ \citenamefont {Pisani}}]{Halo2009a}%
  \BibitemOpen
  \bibfield  {author} {\bibinfo {author} {\bibfnamefont {M.}~\bibnamefont
  {Halo}}, \bibinfo {author} {\bibfnamefont {S.}~\bibnamefont {Casassa}},
  \bibinfo {author} {\bibfnamefont {L.}~\bibnamefont {Maschio}}, \ and\
  \bibinfo {author} {\bibfnamefont {C.}~\bibnamefont {Pisani}},\ }\href
  {\doibase 10.1039/b812870g} {\bibfield  {journal} {\bibinfo  {journal} {Phys.
  Chem. Chem. Phys.}\ }\textbf {\bibinfo {volume} {11}},\ \bibinfo {pages}
  {586} (\bibinfo {year} {2009}{\natexlab{b}})}\BibitemShut {NoStop}%
\bibitem [{\citenamefont {Erba}\ \emph {et~al.}(2009)\citenamefont {Erba},
  \citenamefont {Casassa}, \citenamefont {Dovesi}, \citenamefont {Maschio},\
  and\ \citenamefont {Pisani}}]{Erba2009}%
  \BibitemOpen
  \bibfield  {author} {\bibinfo {author} {\bibfnamefont {A.}~\bibnamefont
  {Erba}}, \bibinfo {author} {\bibfnamefont {S.}~\bibnamefont {Casassa}},
  \bibinfo {author} {\bibfnamefont {R.}~\bibnamefont {Dovesi}}, \bibinfo
  {author} {\bibfnamefont {L.}~\bibnamefont {Maschio}}, \ and\ \bibinfo
  {author} {\bibfnamefont {C.}~\bibnamefont {Pisani}},\ }\href {\doibase
  10.1063/1.3076921} {\bibfield  {journal} {\bibinfo  {journal} {J. Chem.
  Phys.}\ }\textbf {\bibinfo {volume} {130}},\ \bibinfo {pages} {0} (\bibinfo
  {year} {2009})}\BibitemShut {NoStop}%
\bibitem [{\citenamefont {Erba}\ \emph {et~al.}(2011)\citenamefont {Erba},
  \citenamefont {Maschio}, \citenamefont {Pisani},\ and\ \citenamefont
  {Casassa}}]{Erba2011}%
  \BibitemOpen
  \bibfield  {author} {\bibinfo {author} {\bibfnamefont {A.}~\bibnamefont
  {Erba}}, \bibinfo {author} {\bibfnamefont {L.}~\bibnamefont {Maschio}},
  \bibinfo {author} {\bibfnamefont {C.}~\bibnamefont {Pisani}}, \ and\ \bibinfo
  {author} {\bibfnamefont {S.}~\bibnamefont {Casassa}},\ }\href {\doibase
  10.1103/PhysRevB.84.012101} {\bibfield  {journal} {\bibinfo  {journal} {Phys.
  Rev. B}\ }\textbf {\bibinfo {volume} {84}},\ \bibinfo {pages} {012101}
  (\bibinfo {year} {2011})}\BibitemShut {NoStop}%
\bibitem [{\citenamefont {Casassa}\ and\ \citenamefont
  {Demichelis}(2012)}]{Casassa2012}%
  \BibitemOpen
  \bibfield  {author} {\bibinfo {author} {\bibfnamefont {S.}~\bibnamefont
  {Casassa}}\ and\ \bibinfo {author} {\bibfnamefont {R.}~\bibnamefont
  {Demichelis}},\ }\href {\doibase 10.1021/jp300419t} {\bibfield  {journal}
  {\bibinfo  {journal} {J. Phys. Chem. C}\ }\textbf {\bibinfo {volume} {116}},\
  \bibinfo {pages} {13313} (\bibinfo {year} {2012})}\BibitemShut {NoStop}%
\bibitem [{\citenamefont {Fabiano}\ \emph {et~al.}(2009)\citenamefont
  {Fabiano}, \citenamefont {Piacenza}, \citenamefont {D'Agostino},\ and\
  \citenamefont {{Della Sala}}}]{Fabiano2009}%
  \BibitemOpen
  \bibfield  {author} {\bibinfo {author} {\bibfnamefont {E.}~\bibnamefont
  {Fabiano}}, \bibinfo {author} {\bibfnamefont {M.}~\bibnamefont {Piacenza}},
  \bibinfo {author} {\bibfnamefont {S.}~\bibnamefont {D'Agostino}}, \ and\
  \bibinfo {author} {\bibfnamefont {F.}~\bibnamefont {{Della Sala}}},\ }\href
  {\doibase 10.1063/1.3271393} {\bibfield  {journal} {\bibinfo  {journal} {J.
  Chem. Phys.}\ }\textbf {\bibinfo {volume} {131}} (\bibinfo {year} {2009}),\
  10.1063/1.3271393}\BibitemShut {NoStop}%
\bibitem [{\citenamefont {Maschio}\ \emph
  {et~al.}(2010{\natexlab{a}})\citenamefont {Maschio}, \citenamefont {Usvyat},\
  and\ \citenamefont {Civalleri}}]{Maschio2010}%
  \BibitemOpen
  \bibfield  {author} {\bibinfo {author} {\bibfnamefont {L.}~\bibnamefont
  {Maschio}}, \bibinfo {author} {\bibfnamefont {D.}~\bibnamefont {Usvyat}}, \
  and\ \bibinfo {author} {\bibfnamefont {B.}~\bibnamefont {Civalleri}},\ }\href
  {\doibase 10.1039/c002580a} {\bibfield  {journal} {\bibinfo  {journal}
  {CrystEngComm}\ }\textbf {\bibinfo {volume} {12}},\ \bibinfo {pages} {2429}
  (\bibinfo {year} {2010}{\natexlab{a}})}\BibitemShut {NoStop}%
\bibitem [{\citenamefont {Maschio}\ \emph
  {et~al.}(2010{\natexlab{b}})\citenamefont {Maschio}, \citenamefont {Usvyat},
  \citenamefont {Sch{\"{u}}tz},\ and\ \citenamefont
  {Civalleri}}]{Maschio2010a}%
  \BibitemOpen
  \bibfield  {author} {\bibinfo {author} {\bibfnamefont {L.}~\bibnamefont
  {Maschio}}, \bibinfo {author} {\bibfnamefont {D.}~\bibnamefont {Usvyat}},
  \bibinfo {author} {\bibfnamefont {M.}~\bibnamefont {Sch{\"{u}}tz}}, \ and\
  \bibinfo {author} {\bibfnamefont {B.}~\bibnamefont {Civalleri}},\ }\href
  {\doibase 10.1063/1.3372800} {\bibfield  {journal} {\bibinfo  {journal} {J.
  Chem. Phys.}\ }\textbf {\bibinfo {volume} {132}},\ \bibinfo {pages} {134706}
  (\bibinfo {year} {2010}{\natexlab{b}})}\BibitemShut {NoStop}%
\bibitem [{\citenamefont {Schwerdtfeger}\ \emph {et~al.}(2010)\citenamefont
  {Schwerdtfeger}, \citenamefont {Assadollahzadeh},\ and\ \citenamefont
  {Hermann}}]{Schwerdtfeger2010}%
  \BibitemOpen
  \bibfield  {author} {\bibinfo {author} {\bibfnamefont {P.}~\bibnamefont
  {Schwerdtfeger}}, \bibinfo {author} {\bibfnamefont {B.}~\bibnamefont
  {Assadollahzadeh}}, \ and\ \bibinfo {author} {\bibfnamefont {A.}~\bibnamefont
  {Hermann}},\ }\href {\doibase 10.1103/PhysRevB.82.205111} {\bibfield
  {journal} {\bibinfo  {journal} {Phys. Rev. B}\ }\textbf {\bibinfo {volume}
  {82}},\ \bibinfo {pages} {205111} (\bibinfo {year} {2010})}\BibitemShut
  {NoStop}%
\bibitem [{\citenamefont {Nanda}\ and\ \citenamefont
  {Beran}(2012)}]{Nanda2012}%
  \BibitemOpen
  \bibfield  {author} {\bibinfo {author} {\bibfnamefont {K.~D.}\ \bibnamefont
  {Nanda}}\ and\ \bibinfo {author} {\bibfnamefont {G.~J.~O.}\ \bibnamefont
  {Beran}},\ }\href {\doibase 10.1063/1.4764063} {\bibfield  {journal}
  {\bibinfo  {journal} {J. Chem. Phys.}\ }\textbf {\bibinfo {volume} {137}}
  (\bibinfo {year} {2012}),\ 10.1063/1.4764063}\BibitemShut {NoStop}%
\bibitem [{\citenamefont {Stodt}\ and\ \citenamefont
  {H{\"{a}}ttig}(2012)}]{Stodt2012}%
  \BibitemOpen
  \bibfield  {author} {\bibinfo {author} {\bibfnamefont {D.}~\bibnamefont
  {Stodt}}\ and\ \bibinfo {author} {\bibfnamefont {C.}~\bibnamefont
  {H{\"{a}}ttig}},\ }\href {\doibase 10.1063/1.4752478} {\bibfield  {journal}
  {\bibinfo  {journal} {J. Chem. Phys.}\ }\textbf {\bibinfo {volume} {137}},\
  \bibinfo {pages} {114705} (\bibinfo {year} {2012})}\BibitemShut {NoStop}%
\bibitem [{\citenamefont {G{\"{o}}ltl}\ \emph {et~al.}(2012)\citenamefont
  {G{\"{o}}ltl}, \citenamefont {Gr{\"{u}}neis}, \citenamefont {Bu{\v{c}}ko},\
  and\ \citenamefont {Hafner}}]{Goltl2012}%
  \BibitemOpen
  \bibfield  {author} {\bibinfo {author} {\bibfnamefont {F.}~\bibnamefont
  {G{\"{o}}ltl}}, \bibinfo {author} {\bibfnamefont {A.}~\bibnamefont
  {Gr{\"{u}}neis}}, \bibinfo {author} {\bibfnamefont {T.}~\bibnamefont
  {Bu{\v{c}}ko}}, \ and\ \bibinfo {author} {\bibfnamefont {J.}~\bibnamefont
  {Hafner}},\ }\href {\doibase 10.1063/1.4750979} {\bibfield  {journal}
  {\bibinfo  {journal} {J. Chem. Phys.}\ }\textbf {\bibinfo {volume} {137}}
  (\bibinfo {year} {2012}),\ 10.1063/1.4750979}\BibitemShut {NoStop}%
\bibitem [{\citenamefont {M{\"{u}}ller}\ and\ \citenamefont
  {Usvyat}(2013)}]{Muller2013}%
  \BibitemOpen
  \bibfield  {author} {\bibinfo {author} {\bibfnamefont {C.}~\bibnamefont
  {M{\"{u}}ller}}\ and\ \bibinfo {author} {\bibfnamefont {D.}~\bibnamefont
  {Usvyat}},\ }\href {\doibase 10.1021/ct400797w} {\bibfield  {journal}
  {\bibinfo  {journal} {J. Chem. Theory Comput.}\ }\textbf {\bibinfo {volume}
  {9}},\ \bibinfo {pages} {5590} (\bibinfo {year} {2013})}\BibitemShut
  {NoStop}%
\bibitem [{\citenamefont {{Del Ben}}\ \emph {et~al.}(2013)\citenamefont {{Del
  Ben}}, \citenamefont {Sch{\"{o}}nherr}, \citenamefont {Hutter},\ and\
  \citenamefont {VandeVondele}}]{DelBen2013}%
  \BibitemOpen
  \bibfield  {author} {\bibinfo {author} {\bibfnamefont {M.}~\bibnamefont {{Del
  Ben}}}, \bibinfo {author} {\bibfnamefont {M.}~\bibnamefont
  {Sch{\"{o}}nherr}}, \bibinfo {author} {\bibfnamefont {J.}~\bibnamefont
  {Hutter}}, \ and\ \bibinfo {author} {\bibfnamefont {J.}~\bibnamefont
  {VandeVondele}},\ }\href {\doibase 10.1021/jz401931f} {\bibfield  {journal}
  {\bibinfo  {journal} {J. Phys. Chem. Lett.}\ }\textbf {\bibinfo {volume}
  {4}},\ \bibinfo {pages} {3753} (\bibinfo {year} {2013})}\BibitemShut
  {NoStop}%
\bibitem [{\citenamefont {{Del Ben}}\ \emph {et~al.}(2014)\citenamefont {{Del
  Ben}}, \citenamefont {Vandevondele},\ and\ \citenamefont
  {Slater}}]{DelBen2014}%
  \BibitemOpen
  \bibfield  {author} {\bibinfo {author} {\bibfnamefont {M.}~\bibnamefont {{Del
  Ben}}}, \bibinfo {author} {\bibfnamefont {J.}~\bibnamefont {Vandevondele}}, \
  and\ \bibinfo {author} {\bibfnamefont {B.}~\bibnamefont {Slater}},\ }\href
  {\doibase 10.1021/jz501985w} {\bibfield  {journal} {\bibinfo  {journal} {J.
  Phys. Chem. Lett.}\ }\textbf {\bibinfo {volume} {5}},\ \bibinfo {pages}
  {4122} (\bibinfo {year} {2014})}\BibitemShut {NoStop}%
\bibitem [{\citenamefont {{Del Ben}}\ \emph {et~al.}(2015)\citenamefont {{Del
  Ben}}, \citenamefont {Hutter},\ and\ \citenamefont
  {VandeVondele}}]{DelBen2015}%
  \BibitemOpen
  \bibfield  {author} {\bibinfo {author} {\bibfnamefont {M.}~\bibnamefont {{Del
  Ben}}}, \bibinfo {author} {\bibfnamefont {J.}~\bibnamefont {Hutter}}, \ and\
  \bibinfo {author} {\bibfnamefont {J.}~\bibnamefont {VandeVondele}},\ }\href
  {\doibase 10.1063/1.4927325} {\bibfield  {journal} {\bibinfo  {journal} {J.
  Chem. Phys.}\ }\textbf {\bibinfo {volume} {143}} (\bibinfo {year} {2015}),\
  10.1063/1.4927325}\BibitemShut {NoStop}%
\bibitem [{\citenamefont {Torabi}\ \emph {et~al.}(2014)\citenamefont {Torabi},
  \citenamefont {Hammerschmidt}, \citenamefont {Voloshina},\ and\ \citenamefont
  {Paulus}}]{Torabi2014}%
  \BibitemOpen
  \bibfield  {author} {\bibinfo {author} {\bibfnamefont {S.}~\bibnamefont
  {Torabi}}, \bibinfo {author} {\bibfnamefont {L.}~\bibnamefont
  {Hammerschmidt}}, \bibinfo {author} {\bibfnamefont {E.}~\bibnamefont
  {Voloshina}}, \ and\ \bibinfo {author} {\bibfnamefont {B.}~\bibnamefont
  {Paulus}},\ }\href {\doibase 10.1002/qua.24695} {\bibfield  {journal}
  {\bibinfo  {journal} {Int. J. Quantum Chem.}\ }\textbf {\bibinfo {volume}
  {114}},\ \bibinfo {pages} {943} (\bibinfo {year} {2014})}\BibitemShut
  {NoStop}%
\bibitem [{\citenamefont {Hammerschmidt}\ \emph {et~al.}(2015)\citenamefont
  {Hammerschmidt}, \citenamefont {Maschio}, \citenamefont {M{\"{u}}ller},\ and\
  \citenamefont {Paulus}}]{Hammerschmidt2015}%
  \BibitemOpen
  \bibfield  {author} {\bibinfo {author} {\bibfnamefont {L.}~\bibnamefont
  {Hammerschmidt}}, \bibinfo {author} {\bibfnamefont {L.}~\bibnamefont
  {Maschio}}, \bibinfo {author} {\bibfnamefont {C.}~\bibnamefont
  {M{\"{u}}ller}}, \ and\ \bibinfo {author} {\bibfnamefont {B.}~\bibnamefont
  {Paulus}},\ }\href {\doibase 10.1021/ct500841b} {\bibfield  {journal}
  {\bibinfo  {journal} {J. Chem. Theory Comput.}\ }\textbf {\bibinfo {volume}
  {11}},\ \bibinfo {pages} {252} (\bibinfo {year} {2015})}\BibitemShut
  {NoStop}%
\bibitem [{\citenamefont {Kaawar}\ \emph {et~al.}(2016)\citenamefont {Kaawar},
  \citenamefont {M{\"{u}}ller},\ and\ \citenamefont {Paulus}}]{Kaawar2016}%
  \BibitemOpen
  \bibfield  {author} {\bibinfo {author} {\bibfnamefont {Z.}~\bibnamefont
  {Kaawar}}, \bibinfo {author} {\bibfnamefont {C.}~\bibnamefont
  {M{\"{u}}ller}}, \ and\ \bibinfo {author} {\bibfnamefont {B.}~\bibnamefont
  {Paulus}},\ }\href {\doibase 10.1016/j.susc.2016.06.021} {\bibfield
  {journal} {\bibinfo  {journal} {Surf. Sci.}\ } (\bibinfo {year} {2016}),\
  10.1016/j.susc.2016.06.021}\BibitemShut {NoStop}%
\bibitem [{\citenamefont {Marsman}\ \emph {et~al.}(2009)\citenamefont
  {Marsman}, \citenamefont {Gr{\"{u}}neis}, \citenamefont {Paier},\ and\
  \citenamefont {Kresse}}]{Marsman2009}%
  \BibitemOpen
  \bibfield  {author} {\bibinfo {author} {\bibfnamefont {M.}~\bibnamefont
  {Marsman}}, \bibinfo {author} {\bibfnamefont {A.}~\bibnamefont
  {Gr{\"{u}}neis}}, \bibinfo {author} {\bibfnamefont {J.}~\bibnamefont
  {Paier}}, \ and\ \bibinfo {author} {\bibfnamefont {G.}~\bibnamefont
  {Kresse}},\ }\href {\doibase 10.1063/1.3126249} {\bibfield  {journal}
  {\bibinfo  {journal} {J. Chem. Phys.}\ }\textbf {\bibinfo {volume} {130}},\
  \bibinfo {pages} {1} (\bibinfo {year} {2009})}\BibitemShut {NoStop}%
\bibitem [{\citenamefont {Gr{\"{u}}neis}\ \emph {et~al.}(2010)\citenamefont
  {Gr{\"{u}}neis}, \citenamefont {Marsman},\ and\ \citenamefont
  {Kresse}}]{Gruneis2010}%
  \BibitemOpen
  \bibfield  {author} {\bibinfo {author} {\bibfnamefont {A.}~\bibnamefont
  {Gr{\"{u}}neis}}, \bibinfo {author} {\bibfnamefont {M.}~\bibnamefont
  {Marsman}}, \ and\ \bibinfo {author} {\bibfnamefont {G.}~\bibnamefont
  {Kresse}},\ }\href {\doibase 10.1063/1.3466765} {\bibfield  {journal}
  {\bibinfo  {journal} {J. Chem. Phys.}\ }\textbf {\bibinfo {volume} {133}},\
  \bibinfo {pages} {74107} (\bibinfo {year} {2010})}\BibitemShut {NoStop}%
\bibitem [{\citenamefont {Pisani}\ \emph {et~al.}(2005)\citenamefont {Pisani},
  \citenamefont {Busso}, \citenamefont {Capecchi}, \citenamefont {Casassa},
  \citenamefont {Dovesi}, \citenamefont {Maschio}, \citenamefont
  {Zicovich-Wilson},\ and\ \citenamefont {Sch{\"{u}}tz}}]{Pisani2005}%
  \BibitemOpen
  \bibfield  {author} {\bibinfo {author} {\bibfnamefont {C.}~\bibnamefont
  {Pisani}}, \bibinfo {author} {\bibfnamefont {M.}~\bibnamefont {Busso}},
  \bibinfo {author} {\bibfnamefont {G.}~\bibnamefont {Capecchi}}, \bibinfo
  {author} {\bibfnamefont {S.}~\bibnamefont {Casassa}}, \bibinfo {author}
  {\bibfnamefont {R.}~\bibnamefont {Dovesi}}, \bibinfo {author} {\bibfnamefont
  {L.}~\bibnamefont {Maschio}}, \bibinfo {author} {\bibfnamefont
  {C.}~\bibnamefont {Zicovich-Wilson}}, \ and\ \bibinfo {author} {\bibfnamefont
  {M.}~\bibnamefont {Sch{\"{u}}tz}},\ }\href {\doibase 10.1063/1.1857479}
  {\bibfield  {journal} {\bibinfo  {journal} {J. Chem. Phys.}\ }\textbf
  {\bibinfo {volume} {122}} (\bibinfo {year} {2005}),\
  10.1063/1.1857479}\BibitemShut {NoStop}%
\bibitem [{\citenamefont {Pisani}\ \emph {et~al.}(2008)\citenamefont {Pisani},
  \citenamefont {Maschio}, \citenamefont {Casassa}, \citenamefont {Halo},
  \citenamefont {Sch{\"{u}}tz},\ and\ \citenamefont {Usvyat}}]{Pisani2008}%
  \BibitemOpen
  \bibfield  {author} {\bibinfo {author} {\bibfnamefont {C.}~\bibnamefont
  {Pisani}}, \bibinfo {author} {\bibfnamefont {L.}~\bibnamefont {Maschio}},
  \bibinfo {author} {\bibfnamefont {S.}~\bibnamefont {Casassa}}, \bibinfo
  {author} {\bibfnamefont {M.}~\bibnamefont {Halo}}, \bibinfo {author}
  {\bibfnamefont {M.}~\bibnamefont {Sch{\"{u}}tz}}, \ and\ \bibinfo {author}
  {\bibfnamefont {D.}~\bibnamefont {Usvyat}},\ }\href {\doibase
  10.1002/jcc.20975} {\bibfield  {journal} {\bibinfo  {journal} {J. Comput.
  Chem.}\ }\textbf {\bibinfo {volume} {29}},\ \bibinfo {pages} {2113} (\bibinfo
  {year} {2008})},\ \Eprint {http://arxiv.org/abs/NIHMS150003}
  {arXiv:NIHMS150003} \BibitemShut {NoStop}%
\bibitem [{\citenamefont {Usvyat}\ \emph {et~al.}(2015)\citenamefont {Usvyat},
  \citenamefont {Maschio},\ and\ \citenamefont {Sch{\"{u}}tz}}]{Usvyat2015}%
  \BibitemOpen
  \bibfield  {author} {\bibinfo {author} {\bibfnamefont {D.}~\bibnamefont
  {Usvyat}}, \bibinfo {author} {\bibfnamefont {L.}~\bibnamefont {Maschio}}, \
  and\ \bibinfo {author} {\bibfnamefont {M.}~\bibnamefont {Sch{\"{u}}tz}},\
  }\href {\doibase 10.1063/1.4921301} {\bibfield  {journal} {\bibinfo
  {journal} {J. Chem. Phys.}\ }\textbf {\bibinfo {volume} {143}} (\bibinfo
  {year} {2015}),\ 10.1063/1.4921301}\BibitemShut {NoStop}%
\bibitem [{\citenamefont {Katouda}\ and\ \citenamefont
  {Nagase}(2010)}]{Katouda2010}%
  \BibitemOpen
  \bibfield  {author} {\bibinfo {author} {\bibfnamefont {M.}~\bibnamefont
  {Katouda}}\ and\ \bibinfo {author} {\bibfnamefont {S.}~\bibnamefont
  {Nagase}},\ }\href {\doibase 10.1063/1.3503153} {\bibfield  {journal}
  {\bibinfo  {journal} {J. Chem. Phys.}\ }\textbf {\bibinfo {volume} {133}}
  (\bibinfo {year} {2010}),\ 10.1063/1.3503153}\BibitemShut {NoStop}%
\bibitem [{\citenamefont {Izmaylov}\ and\ \citenamefont
  {Scuseria}(2008)}]{Izmaylov2008}%
  \BibitemOpen
  \bibfield  {author} {\bibinfo {author} {\bibfnamefont {A.~F.}\ \bibnamefont
  {Izmaylov}}\ and\ \bibinfo {author} {\bibfnamefont {G.~E.}\ \bibnamefont
  {Scuseria}},\ }\href {\doibase 10.1039/b803274m} {\bibfield  {journal}
  {\bibinfo  {journal} {Phys. Chem. Chem. Phys.}\ }\textbf {\bibinfo {volume}
  {10}},\ \bibinfo {pages} {3421} (\bibinfo {year} {2008})}\BibitemShut
  {NoStop}%
\bibitem [{\citenamefont {Maschio}\ \emph {et~al.}(2007)\citenamefont
  {Maschio}, \citenamefont {Usvyat}, \citenamefont {Manby}, \citenamefont
  {Casassa}, \citenamefont {Pisani},\ and\ \citenamefont
  {Sch{\"{u}}tz}}]{Maschio2007}%
  \BibitemOpen
  \bibfield  {author} {\bibinfo {author} {\bibfnamefont {L.}~\bibnamefont
  {Maschio}}, \bibinfo {author} {\bibfnamefont {D.}~\bibnamefont {Usvyat}},
  \bibinfo {author} {\bibfnamefont {F.~R.}\ \bibnamefont {Manby}}, \bibinfo
  {author} {\bibfnamefont {S.}~\bibnamefont {Casassa}}, \bibinfo {author}
  {\bibfnamefont {C.}~\bibnamefont {Pisani}}, \ and\ \bibinfo {author}
  {\bibfnamefont {M.}~\bibnamefont {Sch{\"{u}}tz}},\ }\href {\doibase
  10.1103/PhysRevB.76.075101} {\bibfield  {journal} {\bibinfo  {journal} {Phys.
  Rev. B - Condens. Matter Mater. Phys.}\ }\textbf {\bibinfo {volume} {76}},\
  \bibinfo {pages} {1} (\bibinfo {year} {2007})}\BibitemShut {NoStop}%
\bibitem [{\citenamefont {Usvyat}\ \emph {et~al.}(2006)\citenamefont {Usvyat},
  \citenamefont {Maschio}, \citenamefont {Manby}, \citenamefont {Casassa},
  \citenamefont {Sch{\"{u}}tz},\ and\ \citenamefont {Pisani}}]{Usvyat2006}%
  \BibitemOpen
  \bibfield  {author} {\bibinfo {author} {\bibfnamefont {D.}~\bibnamefont
  {Usvyat}}, \bibinfo {author} {\bibfnamefont {L.}~\bibnamefont {Maschio}},
  \bibinfo {author} {\bibfnamefont {F.}~\bibnamefont {Manby}}, \bibinfo
  {author} {\bibfnamefont {S.}~\bibnamefont {Casassa}}, \bibinfo {author}
  {\bibfnamefont {M.}~\bibnamefont {Sch{\"{u}}tz}}, \ and\ \bibinfo {author}
  {\bibfnamefont {C.}~\bibnamefont {Pisani}},\ }\href {\doibase
  10.1103/PhysRevB.76.075102} {\bibfield  {journal} {\bibinfo  {journal} {Phys.
  Rev. B}\ }\textbf {\bibinfo {volume} {76}},\ \bibinfo {pages} {75102}
  (\bibinfo {year} {2006})}\BibitemShut {NoStop}%
\bibitem [{\citenamefont {{Del Ben}}\ \emph {et~al.}(2012)\citenamefont {{Del
  Ben}}, \citenamefont {Hutter},\ and\ \citenamefont
  {Vandevondele}}]{DelBen2012}%
  \BibitemOpen
  \bibfield  {author} {\bibinfo {author} {\bibfnamefont {M.}~\bibnamefont {{Del
  Ben}}}, \bibinfo {author} {\bibfnamefont {J.}~\bibnamefont {Hutter}}, \ and\
  \bibinfo {author} {\bibfnamefont {J.}~\bibnamefont {Vandevondele}},\ }\href
  {\doibase 10.1021/ct300531w} {\bibfield  {journal} {\bibinfo  {journal} {J.
  Chem. Theory Comput.}\ }\textbf {\bibinfo {volume} {8}},\ \bibinfo {pages}
  {4177} (\bibinfo {year} {2012})}\BibitemShut {NoStop}%
\bibitem [{\citenamefont {Suhai}(1983)}]{Suhai1983}%
  \BibitemOpen
  \bibfield  {author} {\bibinfo {author} {\bibfnamefont {S.}~\bibnamefont
  {Suhai}},\ }\href {\doibase 10.1103/PhysRevB.27.3506} {\bibfield  {journal}
  {\bibinfo  {journal} {Phys. Rev. B}\ }\textbf {\bibinfo {volume} {27}},\
  \bibinfo {pages} {3506} (\bibinfo {year} {1983})},\ \Eprint
  {http://arxiv.org/abs/1011.1669} {arXiv:1011.1669} \BibitemShut {NoStop}%
\bibitem [{\citenamefont {Sun}\ and\ \citenamefont {Bartlett}(1996)}]{Sun1996}%
  \BibitemOpen
  \bibfield  {author} {\bibinfo {author} {\bibfnamefont {J.}~\bibnamefont
  {Sun}}\ and\ \bibinfo {author} {\bibfnamefont {R.}~\bibnamefont {Bartlett}},\
  }\href {\doibase 10.1063/1.471545} {\bibfield  {journal} {\bibinfo  {journal}
  {J. Chem. Phys.}\ }\textbf {\bibinfo {volume} {104}},\ \bibinfo {pages}
  {8553} (\bibinfo {year} {1996})}\BibitemShut {NoStop}%
\bibitem [{\citenamefont {Willow}\ \emph {et~al.}(2014)\citenamefont {Willow},
  \citenamefont {Kim},\ and\ \citenamefont {Hirata}}]{Willow2014}%
  \BibitemOpen
  \bibfield  {author} {\bibinfo {author} {\bibfnamefont {S.~Y.}\ \bibnamefont
  {Willow}}, \bibinfo {author} {\bibfnamefont {K.~S.}\ \bibnamefont {Kim}}, \
  and\ \bibinfo {author} {\bibfnamefont {S.}~\bibnamefont {Hirata}},\ }\href
  {\doibase 10.1103/PhysRevB.90.201110} {\bibfield  {journal} {\bibinfo
  {journal} {Phys. Rev. B - Condens. Matter Mater. Phys.}\ }\textbf {\bibinfo
  {volume} {90}},\ \bibinfo {pages} {1} (\bibinfo {year} {2014})}\BibitemShut
  {NoStop}%
\bibitem [{\citenamefont {Neuhauser}\ \emph {et~al.}(2013)\citenamefont
  {Neuhauser}, \citenamefont {Rabani},\ and\ \citenamefont
  {Baer}}]{Neuhauser2013}%
  \BibitemOpen
  \bibfield  {author} {\bibinfo {author} {\bibfnamefont {D.}~\bibnamefont
  {Neuhauser}}, \bibinfo {author} {\bibfnamefont {E.}~\bibnamefont {Rabani}}, \
  and\ \bibinfo {author} {\bibfnamefont {R.}~\bibnamefont {Baer}},\ }\href
  {\doibase 10.1021/ct300946j} {\bibfield  {journal} {\bibinfo  {journal} {J.
  Chem. Theory Comput.}\ }\textbf {\bibinfo {volume} {9}},\ \bibinfo {pages}
  {24} (\bibinfo {year} {2013})}\BibitemShut {NoStop}%
\bibitem [{\citenamefont {Ge}\ \emph {et~al.}(2014)\citenamefont {Ge},
  \citenamefont {Gao}, \citenamefont {Baer}, \citenamefont {Rabani},\ and\
  \citenamefont {Neuhauser}}]{Ge2014}%
  \BibitemOpen
  \bibfield  {author} {\bibinfo {author} {\bibfnamefont {Q.}~\bibnamefont
  {Ge}}, \bibinfo {author} {\bibfnamefont {Y.}~\bibnamefont {Gao}}, \bibinfo
  {author} {\bibfnamefont {R.}~\bibnamefont {Baer}}, \bibinfo {author}
  {\bibfnamefont {E.}~\bibnamefont {Rabani}}, \ and\ \bibinfo {author}
  {\bibfnamefont {D.}~\bibnamefont {Neuhauser}},\ }\href {\doibase
  10.1021/jz402206m} {\bibfield  {journal} {\bibinfo  {journal} {J. Phys. Chem.
  Lett.}\ }\textbf {\bibinfo {volume} {5}},\ \bibinfo {pages} {185} (\bibinfo
  {year} {2014})}\BibitemShut {NoStop}%
\bibitem [{\citenamefont {Takeshita}\ \emph {et~al.}(2017)\citenamefont
  {Takeshita}, \citenamefont {Jong}, \citenamefont {Neuhauser}, \citenamefont
  {Baer},\ and\ \citenamefont {Rabani}}]{Takeshita}%
  \BibitemOpen
  \bibfield  {author} {\bibinfo {author} {\bibfnamefont {T.~Y.}\ \bibnamefont
  {Takeshita}}, \bibinfo {author} {\bibfnamefont {W.~A.~D.}\ \bibnamefont
  {Jong}}, \bibinfo {author} {\bibfnamefont {D.}~\bibnamefont {Neuhauser}},
  \bibinfo {author} {\bibfnamefont {R.}~\bibnamefont {Baer}}, \ and\ \bibinfo
  {author} {\bibfnamefont {E.}~\bibnamefont {Rabani}},\ }\href {\doibase
  10.1021/acs.jctc.7b00343} {\bibfield  {journal} {\bibinfo  {journal} {J.
  Chem. Theory Comput.}\ }\textbf {\bibinfo {volume} {13}},\ \bibinfo {pages}
  {4605} (\bibinfo {year} {2017})}\BibitemShut {NoStop}%
\bibitem [{\citenamefont {Neuhauser}\ \emph {et~al.}(2017)\citenamefont
  {Neuhauser}, \citenamefont {Baer},\ and\ \citenamefont
  {Zgid}}]{Neuhauser2017}%
  \BibitemOpen
  \bibfield  {author} {\bibinfo {author} {\bibfnamefont {D.}~\bibnamefont
  {Neuhauser}}, \bibinfo {author} {\bibfnamefont {R.}~\bibnamefont {Baer}}, \
  and\ \bibinfo {author} {\bibfnamefont {D.}~\bibnamefont {Zgid}},\ }\href
  {\doibase 10.1021/acs.jctc.7b00792} {\bibfield  {journal} {\bibinfo
  {journal} {J. Chem. Theory Comput.}\ }\textbf {\bibinfo {volume} {13}},\
  \bibinfo {pages} {5396} (\bibinfo {year} {2017})},\ \Eprint
  {http://arxiv.org/abs/1707.08296} {arXiv:1707.08296} \BibitemShut {NoStop}%
\bibitem [{\citenamefont {Kresse}\ and\ \citenamefont
  {Hafner}(1993)}]{Kresse1993}%
  \BibitemOpen
  \bibfield  {author} {\bibinfo {author} {\bibfnamefont {G.}~\bibnamefont
  {Kresse}}\ and\ \bibinfo {author} {\bibfnamefont {J.}~\bibnamefont
  {Hafner}},\ }\href {\doibase 10.1103/PhysRevB.47.558} {\bibfield  {journal}
  {\bibinfo  {journal} {Phys. Rev. B}\ }\textbf {\bibinfo {volume} {47}},\
  \bibinfo {pages} {558} (\bibinfo {year} {1993})},\ \Eprint
  {http://arxiv.org/abs/0927-0256(96)00008} {arXiv:0927-0256(96)00008
  [10.1016]} \BibitemShut {NoStop}%
\bibitem [{\citenamefont {Kresse}\ and\ \citenamefont
  {Joubert}(1999)}]{Kresse1999}%
  \BibitemOpen
  \bibfield  {author} {\bibinfo {author} {\bibfnamefont {G.}~\bibnamefont
  {Kresse}}\ and\ \bibinfo {author} {\bibfnamefont {D.}~\bibnamefont
  {Joubert}},\ }\href {\doibase 10.1103/PhysRevB.59.1758} {\bibfield  {journal}
  {\bibinfo  {journal} {Phys. Rev. B}\ }\textbf {\bibinfo {volume} {59}},\
  \bibinfo {pages} {1758} (\bibinfo {year} {1999})}\BibitemShut {NoStop}%
\bibitem [{\citenamefont {H{\"{a}}ser}\ and\ \citenamefont
  {Alml{\"{o}}f}(1992)}]{Haser1992}%
  \BibitemOpen
  \bibfield  {author} {\bibinfo {author} {\bibfnamefont {M.}~\bibnamefont
  {H{\"{a}}ser}}\ and\ \bibinfo {author} {\bibfnamefont {J.}~\bibnamefont
  {Alml{\"{o}}f}},\ }\href {\doibase 10.1063/1.462485} {\bibfield  {journal}
  {\bibinfo  {journal} {J. Chem. Phys.}\ }\textbf {\bibinfo {volume} {96}},\
  \bibinfo {pages} {489} (\bibinfo {year} {1992})}\BibitemShut {NoStop}%
\bibitem [{\citenamefont {Kaltak}\ \emph {et~al.}(2014)\citenamefont {Kaltak},
  \citenamefont {Klime{\v{s}}},\ and\ \citenamefont {Kresse}}]{Kaltak2014}%
  \BibitemOpen
  \bibfield  {author} {\bibinfo {author} {\bibfnamefont {M.}~\bibnamefont
  {Kaltak}}, \bibinfo {author} {\bibfnamefont {J.}~\bibnamefont
  {Klime{\v{s}}}}, \ and\ \bibinfo {author} {\bibfnamefont {G.}~\bibnamefont
  {Kresse}},\ }\href {\doibase 10.1021/ct5001268} {\bibfield  {journal}
  {\bibinfo  {journal} {J. Chem. Theory Comput.}\ }\textbf {\bibinfo {volume}
  {10}},\ \bibinfo {pages} {2498} (\bibinfo {year} {2014})}\BibitemShut
  {NoStop}%
\bibitem [{\citenamefont {Welford}(1962)}]{Welford1962}%
  \BibitemOpen
  \bibfield  {author} {\bibinfo {author} {\bibfnamefont {B.~P.}\ \bibnamefont
  {Welford}},\ }\href {\doibase 10.2307/1266577} {\bibfield  {journal}
  {\bibinfo  {journal} {Technometrics}\ }\textbf {\bibinfo {volume} {4}},\
  \bibinfo {pages} {419} (\bibinfo {year} {1962})}\BibitemShut {NoStop}%
\bibitem [{\citenamefont {Willow}\ \emph {et~al.}(2013)\citenamefont {Willow},
  \citenamefont {Hermes}, \citenamefont {Kim},\ and\ \citenamefont
  {Hirata}}]{Willow2013}%
  \BibitemOpen
  \bibfield  {author} {\bibinfo {author} {\bibfnamefont {S.~Y.}\ \bibnamefont
  {Willow}}, \bibinfo {author} {\bibfnamefont {M.~R.}\ \bibnamefont {Hermes}},
  \bibinfo {author} {\bibfnamefont {K.~S.}\ \bibnamefont {Kim}}, \ and\
  \bibinfo {author} {\bibfnamefont {S.}~\bibnamefont {Hirata}},\ }\href
  {\doibase 10.1021/ct400557z} {\bibfield  {journal} {\bibinfo  {journal} {J.
  Chem. Theory Comput.}\ }\textbf {\bibinfo {volume} {9}},\ \bibinfo {pages}
  {4396} (\bibinfo {year} {2013})}\BibitemShut {NoStop}%
\bibitem [{\citenamefont {Harl}\ and\ \citenamefont {Kresse}(2008)}]{Harl2008}%
  \BibitemOpen
  \bibfield  {author} {\bibinfo {author} {\bibfnamefont {J.}~\bibnamefont
  {Harl}}\ and\ \bibinfo {author} {\bibfnamefont {G.}~\bibnamefont {Kresse}},\
  }\href {\doibase 10.1103/PhysRevB.77.045136} {\bibfield  {journal} {\bibinfo
  {journal} {Phys. Rev. B - Condens. Matter Mater. Phys.}\ }\textbf {\bibinfo
  {volume} {77}},\ \bibinfo {pages} {1} (\bibinfo {year} {2008})}\BibitemShut
  {NoStop}%
\bibitem [{\citenamefont {Shepherd}\ \emph {et~al.}(2012)\citenamefont
  {Shepherd}, \citenamefont {Gr{\"{u}}neis}, \citenamefont {Booth},
  \citenamefont {Kresse},\ and\ \citenamefont {Alavi}}]{Shepherd2012}%
  \BibitemOpen
  \bibfield  {author} {\bibinfo {author} {\bibfnamefont {J.~J.}\ \bibnamefont
  {Shepherd}}, \bibinfo {author} {\bibfnamefont {A.}~\bibnamefont
  {Gr{\"{u}}neis}}, \bibinfo {author} {\bibfnamefont {G.~H.}\ \bibnamefont
  {Booth}}, \bibinfo {author} {\bibfnamefont {G.}~\bibnamefont {Kresse}}, \
  and\ \bibinfo {author} {\bibfnamefont {A.}~\bibnamefont {Alavi}},\ }\href
  {\doibase 10.1103/PhysRevB.86.035111} {\bibfield  {journal} {\bibinfo
  {journal} {Phys. Rev. B - Condens. Matter Mater. Phys.}\ }\textbf {\bibinfo
  {volume} {86}},\ \bibinfo {pages} {1} (\bibinfo {year} {2012})},\ \Eprint
  {http://arxiv.org/abs/1202.4990} {arXiv:1202.4990} \BibitemShut {NoStop}%
\bibitem [{\citenamefont {Klime{\v{s}}}\ \emph {et~al.}(2010)\citenamefont
  {Klime{\v{s}}}, \citenamefont {Bowler},\ and\ \citenamefont
  {Michaelides}}]{Klimes2010}%
  \BibitemOpen
  \bibfield  {author} {\bibinfo {author} {\bibfnamefont {J.}~\bibnamefont
  {Klime{\v{s}}}}, \bibinfo {author} {\bibfnamefont {D.~R.}\ \bibnamefont
  {Bowler}}, \ and\ \bibinfo {author} {\bibfnamefont {A.}~\bibnamefont
  {Michaelides}},\ }\href {\doibase 10.1088/0953-8984/22/2/022201} {\bibfield
  {journal} {\bibinfo  {journal} {J. Phys. Condens. Matter}\ }\textbf {\bibinfo
  {volume} {22}},\ \bibinfo {pages} {022201} (\bibinfo {year} {2010})},\
  \Eprint {http://arxiv.org/abs/0910.0438} {arXiv:0910.0438} \BibitemShut
  {NoStop}%
\end{thebibliography}%

\end{document}